\documentclass[showpacs,preprintnumbers,amsmath,amssymb,superscriptaddress]{revtex4}

\usepackage{calc}
\usepackage{graphicx,color}
\usepackage{bm}
\usepackage{color}

\newcommand{\pa}{\partial}
\newcommand{\pd}[2]{\frac{\partial #1}{\partial #2}}
\newcommand{\pdds}[2]{\frac{\partial^2 #1}{\partial #2^2}}

\newcommand{\nm}[1]{\left|#1\right|} 
\newcommand{\tp}[1]{\tilde{\theta}} 

\def\Ei{\mathrm{Ei}}
\def\Z{\mathbb{Z}}

\def\P{\mathbb{P}}

\begin{document}

\title{Kinetics of active surface-mediated diffusion in spherically symmetric domains}

\author{J.-F. Rupprecht}
\affiliation{Laboratoire de Physique Th\'eorique de la Mati\`ere Condens\'ee
(UMR 7600), case courrier 121, Universit\'e Paris 6, 4 Place Jussieu, 75255
Paris Cedex}

\author{O. B\'enichou}
\affiliation{Laboratoire de Physique Th\'eorique de la Mati\`ere Condens\'ee
(UMR 7600), case courrier 121, Universit\'e Paris 6, 4 Place Jussieu, 75255
Paris Cedex}

\author{D. S. Grebenkov}
\affiliation{Laboratoire de Physique de la Mati\`ere Condens\'ee (UMR 7643),
CNRS -- Ecole Polytechnique, F-91128 Palaiseau Cedex France}

\author{R. Voituriez}
\affiliation{Laboratoire de Physique Th\'eorique de la Mati\`ere Condens\'ee
(UMR 7600), case courrier 121, Universit\'e Paris 6, 4 Place Jussieu, 75255
Paris Cedex}

\date{\today}

\begin{abstract}
We present an exact calculation of the mean first-passage time to a
target on the surface of a 2D or 3D spherical domain, for a molecule
alternating phases of surface diffusion on the domain boundary and
phases of bulk diffusion.  We generalize the results of
\cite{Benichou:2011a} and consider a biased diffusion in a general
annulus with an arbitrary number of regularly spaced targets on a
partially reflecting surface.  The presented approach is based on an
integral equation which can be solved analytically.  Numerically
validated approximation schemes, which provide more tractable
expressions of the mean first-passage time are also proposed.  In the
framework of this minimal model of surface-mediated reactions, we show
analytically that the mean reaction time can be minimized as a
function of the desorption rate from the surface.
\end{abstract}


\maketitle

\section{Introduction}

Reaction kinetics in confined systems where a small number of reactants are involved, such as porous catalysts and
living cells, 
can be limited by the time needed for
molecules to meet and react \cite{Rice:1985,Hanggi:1990a}.  This time is known in random walk theory
as a first-passage time (FPT) \cite{Redner:2001a,Moreau:2003xe,Condamin:2007zl,BenichouO.:2010a}.
%
%
For the specific case of biochemical reactions in living cells, these
general considerations have to incorporate two important features.
First, while passive diffusion is the dominant mode of transport in
chemical systems, active transport has been shown to play a prominent
role in living cells \cite{Alberts:2002}.  As a matter of fact,
various motor proteins such as kinesin and myosin are able to convert
the chemical fuel provided by ATP into mechanical work by interacting
with the filaments of the cytoskeleton.  Many
macromolecules or larger cellular organelles such as vesicles,
lysosomes or mitochondria, can randomly bind and unbind to these
motors \cite{Huet:2006,Loverdo:2008,Arcizet:2008}.  As a result, the
overall transport of such tracers in the cell can be described in a
first approximation as diffusion in a force field
\cite{Lagache:2008a}.
Second, reactions in confined domains like cells generally involve
surface-mediated diffusion that combines bulk transport and surface
diffusion due to non-specific interactions with the domain boundary
(e.g. cell membrane)
\cite{Astumian:1985,bond,Berg:1981,Adam:1968,Sano:1981,Schuss:2007}.
Such two-state paths and the corresponding first-passage properties
have been studied in the broader context of intermittent search
strategies
\cite{Benichou:2005e,Benichou:2005f,Obenichou:2008,Benichou:2011b}
under the hypothesis that the times spent in each state (surface and
bulk) are controlled by an internal clock independent of any
geometrical parameter.  In most cases, the sojourn times in each state
have been assumed to be exponentially distributed
\cite{Obenichou:2008}, with the notable exception of L\'evy
\cite{Lomholt:2008} and deterministic laws
\cite{Benichou:2007,Oshanin:2007a}.  However, in the case of
interfacial reactions, for which molecules react on target sites
located on the surface of the confining domain, the time spent in a
bulk excursion is controlled by the statistics of return to the
surface and therefore by the geometry of the confining domain
\cite{Majumdar:1999,Benichou:2005a,Levitz:2006,Levitz:2008,Chechkin:2009}.
Hence this return time is not an external parameter but is generated
by the very dynamics of the diffusing molecule in confinement.

Recently, such coupling of the intermittent dynamics to the geometry
of the confinement has been explicitly taken into account
\cite{Revelli:2005,Oshanin:2010,Benichou:2010,Benichou:2011a,Rojo:2011,berezhkovskii:054115}.
Exact calculations of the mean FPT to a target on the surface of a 2D
or 3D spherical domain, for a molecule performing surface-mediated
diffusion, have been proposed \cite{Benichou:2010,Benichou:2011a}.
However, these works have been limited to  passive transport alone.
The present article develops a general theoretical framework which in particular
allows one to incorporate the effect of active transport
on surface-mediated diffusion.

More precisely, we extend the results of
\cite{Benichou:2010,Benichou:2011a} in four directions: (i) we
consider the general case of an {\it imperfect adsorption step}, so
that the molecule can bounce several times before being adsorbed to
the confining surface
\cite{Zweifel:1973, Wio:1993, Barzykin:1993,Sapoval:1994,Benichou:2000it,Grebenkov:2006,Singer:2008,Grebenkov:2010a,Grebenkov:2010b};
(ii) the geometry adopted is a {\it general annulus}, whose either
interior, or exterior boundary is purely reflecting; (iii) we take
into account the effect of an {\it exterior radial force field} which,
for instance, can schematically mimic the effect of active transport;
(iv) we consider the possibility of having an {\it arbitrary number of
regularly spaced targets} on the surface.  Relying on an integral
equation approach, we provide an exact solution for the mean FPT, both
for 2D an 3D spherical domains, and for any spherical target size.  We
also develop approximation schemes, numerically validated, that
provide more tractable expressions of the mean first passage time
(MFPT).

The article is organized as follows.  In Sec. \ref{sec:model}, we
define the model under study; in Sec. \ref{sec:general}, we show that
the MFPT can be determined by solving coupled partial differential
equations that can actually be converted into a single integral
equation.  We then provide an exact solution of this integral
equation, as well as an approximate, more tractable, solution.  In
Sec. \ref{sec:particular}, we give fully explicit expressions of the
MFPT by applying this general formalism to particular cases,
representative of the four aforementioned extensions.

\section{The model}
\label{sec:model}

\begin{figure}[h!]
	\centering
    \includegraphics[width=0.8\textwidth]{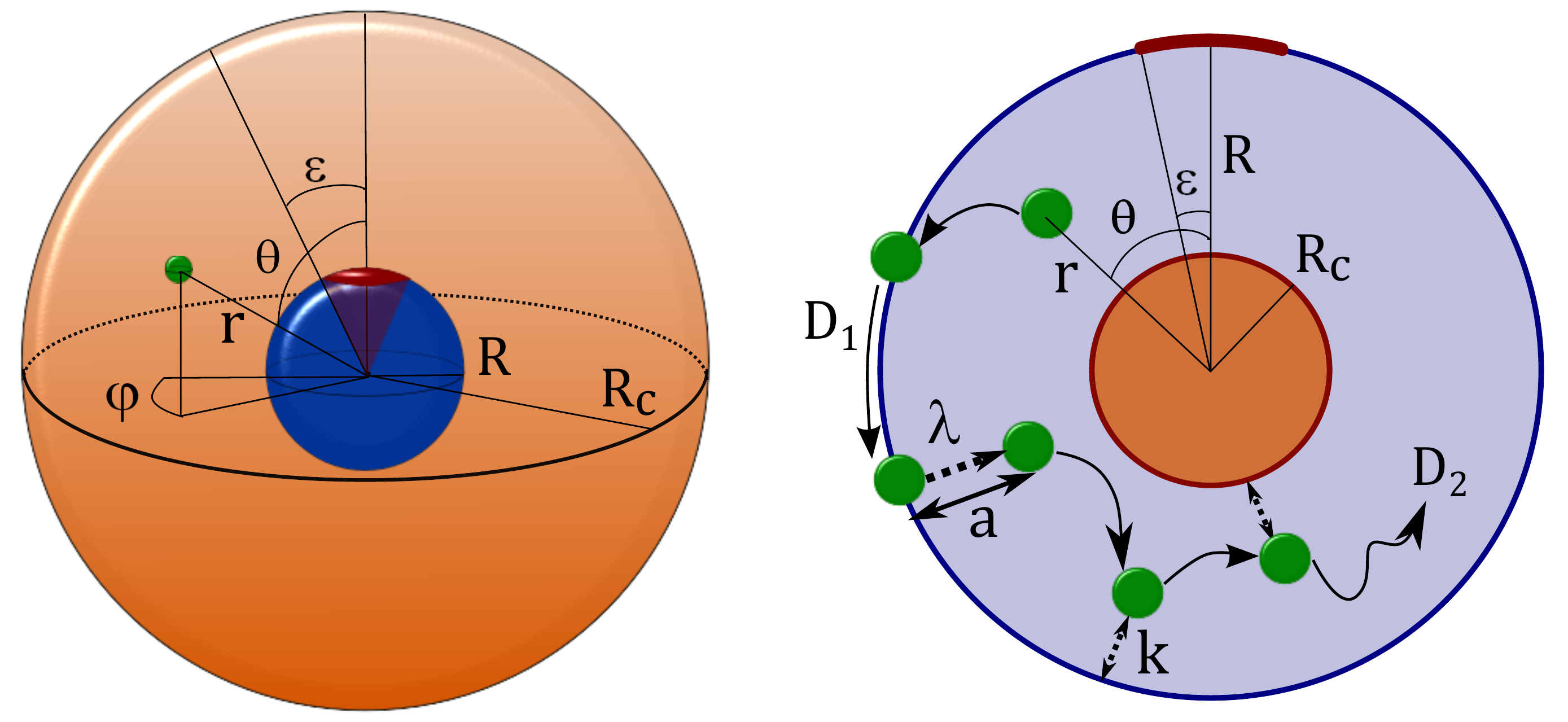}
\caption{
Model - \textit{Left}: Static picture of the entrance problem in 3D -
\textit{Right}: Dynamic picture of the exit problem in 2D.  The green
sphere stands for the diffusing molecule and the red sector stands for
the target.}  
\label{fig:1}
\end{figure}

The surface-mediated process under study is illustrated in
Fig. \ref{fig:1}.  We consider a molecule diffusing in the volume $S$
between two concentric spheres of radii $R$ and $R_{c}$.  The molecule
alternates phases of bulk diffusion (with diffusion coefficient $D_2$)
and phases of surface diffusion on the boundary of the sphere of
radius $R$ (with diffusion coefficient $D_1$) which contains a target.
The target is defined in 2D by the arc
$\theta\in[-\epsilon,\epsilon]$, and in 3D by the region of the sphere
such that $\theta\in[0,\epsilon]$ where $\theta$ is in this case the
elevation angle in spherical coordinates.  Note that as soon as
$\epsilon\neq 0$, the target can be reached both by surface and bulk
diffusion.

In the following, the case $R>R_{c}$ will be called an exit problem
and the case $R_{c}>R$ an entrance problem (Fig. \ref{fig:1}). 
In 3D, the entrance problem can account for the time needed for a
virus initially in the cell (the sphere of radius $R_{c}$) to get into
the nucleus (the sphere of radius $R$) through a single nuclear pore
(the target) in the presence of diffusion on the nuclear membrane.  In
turn, the exit problem in 3D may describe macromolecules searching an
exit from the cell through a channel (or channels) in the cellular
membrane. In that case, the surface of the nucleus is considered as
purely reflecting.  The 2D geometry could correspond to cells that are
confined, as realized in vitro for example in \cite{Hawkins:2009}.

In this model, a molecule hitting the sphere of radius $R_{c}$ is
immediately reflected. In contrast, when the molecule reaches the
sphere of radius $R$, which contains the target, it is imperfectly
adsorbed: the molecule hitting  the boundary at $\bm{r}=(R,\theta)$, $\theta\in[0,\pi]$
is at random either adsorbed to the
sphere of radius $R$ or reflected back in the bulk. The quantity $k$
which describes the rate of adsorption is more precisely defined
through the radiative boundary condition Eq. (\ref{eq:adsorption})
(see also Eq. (\ref{eq:kdefinition}) of the discrete lattice approach
discussed in Appendix \ref{sec:boundary}).  In particular, $k =
\infty$ corresponds to a perfectly adsorbing boundary and $k = 0$ to a
perfectly reflecting boundary.  Notice that for finite $k$, molecules hitting the target from the bulk
can be reflected.

The time spent during each surface exploration on the sphere of radius
$R$ is assumed to follow an exponential law with desorption rate
$\lambda$.  At each desorption event, the molecule is assumed to be
ejected from the surface point ${\bf{r}}=(R,\theta)$ to the bulk point
${\bf{r}}=(R-a,\theta)$.  In what follows, $a$ can be positive or
negative: $a>0$ for the exit problem ($R > R_c$), and $a<0$ for the
entrance problem ($R < R_c$).  Although formulated for any value of
the parameter $a$ such that $\nm{a} \leq \nm{R-R_c}$ (to ensure that
the particle remains inside the domain after reflection), in most
physical situations of interest $\nm{a}$ is much smaller than $R$.
Note finally that a non zero ejection distance $a$ is required in the
limit of perfect adsorption $k = \infty$, otherwise the diffusing molecule would
be instantaneously re-adsorbed on the surface.

\section{General solution}
\label{sec:general}

\subsection{Basic equations}

For the process defined above, the mean first-passage time (MFPT)
satisfies the following backward equations
\begin{eqnarray}  
\label{eq:temps1}
\frac{D_{1}}{R^2} \Delta_{\theta} t_{1}(\theta) + 
\lambda  \left(t_{2}(R-a,\theta) - t_{1}(\theta) \right) &=& - 1  \quad (\epsilon < \theta < \pi), \\
\label{eq:temps2} 
D_{2} \left( \Delta_{r} + \frac{v(r)}{D_{2}} \; \pa_{r} + \frac{\Delta_{\theta}}{r^2}  \right) \; t_{2}(r,\theta) &=& -1 \quad
 ((r,\theta) \in S),
\end{eqnarray}
where: (i) $t_1(\theta)$ stands for the MFPT for a molecule initially
on the sphere of radius $R$ at angle $\theta$, and $t_2(r,\theta)$
stands for the MFPT for a molecule initially at a bulk point
$(r,\theta)$ within the annulus $S = (R_{c},R)\times[0,\pi]$; note
that, due to the symmetry $t_i(\theta) = t_i(-\theta)$, in 2D $\theta$
can be restricted to $[0,\pi]$; (ii) the radial and angular Laplace
operators are respectively
\begin{equation*}
\Delta_{r} = \pdds{}{r} + \frac{d-1}{r} \pd{}{r} , \qquad 
\Delta_{\theta} = (\sin \theta)^{2-d} \; \pa_{\theta}\; (\sin\theta)^{d-2} \; \pa_{\theta},
\end{equation*}
and $d$ stands for the space dimension (in practice, $d$ will be taken
equal to 2 or 3); (iii) $v(r)$ is the radial velocity of the molecule
resulting from an external force.

In Eqs. (\ref{eq:temps1}, \ref{eq:temps2}), the first terms of the
left hand side account for diffusion respectively on the surface and
in the bulk, while the second term of Eq. (\ref{eq:temps1}) describes
desorption events.  These equations have to be completed by boundary
conditions:

(i) reflecting boundary condition on the sphere of radius $R_c \geq 0$
\begin{equation} 
\label{eq:bcrefl}
\pd{t_{2}}{r}_{\lvert {\bf{r}}=(R_c,\theta)} =0  \qquad (0\leq \theta \leq \pi)
\end{equation}
(note that this condition holds even in the presence of the velocity
field $v(r)$, see e.g. \cite{Benichou:2009});

(ii) radiative boundary condition
\begin{eqnarray}
\label{eq:adsorption}
\pd{t_{2}}{r}_{\lvert {\bf{r}}=(R,\theta)} =  k \{ t_{1}(\theta) - t_{2}(R,\theta) \} \qquad (0\leq \theta \leq \pi),
\end{eqnarray}
which describes the partial adsorption events on the sphere of radius $R$
(see Appendix \ref{sec:boundary} for justification of this boundary
condition).  For the exit problem ($R>R_{c}$), the radial axis points
towards the surface and $k>0$, while for the entrance problem
($R<R_{c}$), the radial axis points outwards the surface and $k<0$.
Finally, the limit $k = \pm \infty$ describes the perfect adsorption
for which the above condition reads as $t_1(\theta) = t_2(R,\theta)$.

(iii) Dirichlet boundary condition
\begin{equation} 
\label{eq:bcadsorb} 
t_1(\theta) = 0    \qquad (0 \leq \theta \leq \epsilon),
\end{equation}
which expresses that the target is an absorbing zone (the search process is
stopped on the target).

In what follows we will use two dimensionless quantities
\begin{eqnarray}
  x\equiv 1-a/R, \\
  \omega \equiv R\sqrt{\lambda/D_{1}} ,
\end{eqnarray}
and the operator $L$ acting on a function $f$ as
\begin{equation} 
\label{def:l}
(L f)(r) \equiv f(r-a) - f(r)-  \frac{1}{k}~ \pa_{r} f(r).
\end{equation}

\subsection{General integral equation}

We generalize the approach presented in \cite{Benichou:2011a} and show
that the coupled Eqs. (\ref{eq:temps1}, \ref{eq:temps2}) with the
boundary conditions (\ref{eq:bcrefl}-\ref{eq:bcadsorb}) lead to the
integral equation (\ref{eq:diffgenerale}) for $t_1$ only.

The starting point is a Fourier decomposition of
$t_2$. Eq. (\ref{eq:temps2}) is easily shown to be satisfied by
\begin{equation} 
\label{eq:t2_def} 
t_{2}(r,\theta) = \alpha_{0}+ \frac{1}{D_{2}} \hat{f}(r) + \beta_{0} \; f_{0}(r) +
\sum^{\infty}_{n=1} \alpha_{n} f_{n}(r) V_{n}(\theta ) +
\sum^{\infty}_{n=1} \alpha_{-n} f_{-n}(r) V_{n}(\theta ),
\end{equation} 
with coefficients $\alpha_{n}$ to be determined, and

(i) $\hat{f}(r)$ is a rotation-invariant solution of
Eq. (\ref{eq:temps2}) regular at $r=0$, i.e.
\begin{equation}
\label{eq:fhat}
\left(\Delta_r + \frac{v(r)}{D_2}\; \pa_r\right) \hat{f}(r) = - 1 ,
\end{equation}
the choice of $\hat{f}(r)$ being up to an additive constant;

(ii) $f_{0}(r)$ is a non-constant solution of the homogeneous equation
\begin{equation}
\label{eq:f0}
\left(\Delta_r + \frac{v(r)}{D_2}\; \pa_r\right) f_0(r) = 0 ,
\end{equation}
the choice of $f_0(r)$ being up to an additive constant and a
multiplicative prefactor.  It can be shown that $f_{0}(r)$ necessarily
diverges at $r=0$ in our cases of interest;

(iii) the set of functions $\{f_{n}(r) , V_{n}(\theta)\}_{n\in
\mathbb{Z}}$ is an eigenbasis of the homogeneous equation
associated to Eq. (\ref{eq:temps2}):
\begin{eqnarray}
\label{eq:Vn_def}  
- \Delta_{\theta} V_{n}(\theta ) &=& \rho_{n} V_{n}(\theta )   \quad  (n\geq 0),  \\
\label{eq:fn_def}   
r^2 \left( \Delta_{r} + \frac{v(r)}{D_{2}} \; \pa_{r} \right) f_{n}(r) &=& \rho_{|n|} f_{n}(r)   \quad (n\in \mathbb{Z}) , 
\end{eqnarray}
with $V_{-n}(\theta) = V_n(\theta)$ due to the reflection symmetry,
and
\begin{equation}
\label{eq:rho}
\rho_{n} = \begin{cases} n^2  \hskip 13mm (d = 2) , \cr   n(n+1)  \quad (d = 3). \end{cases}
\end{equation}
We set 
\begin{equation}
V_{n}(\theta) = \begin{cases}  \begin{cases} 1  \hskip 18mm  (n = 0)  \cr
\sqrt{2} \cos(n\theta)  \quad (n > 0) \end{cases}  \qquad (d = 2) , \cr 
\sqrt{2n+1}~P_{n}(\cos \theta) \quad (n\geq 0) \quad (d = 3),  \end{cases}
\end{equation}
where $P_n(z)$ are Legendre polynomials.  In turn, the functions
$f_n(r)$ which depend on the velocity field $v(r)$, will be determined
individually case by case (see Sec. \ref{sec:particular}).

In the following, we will use two inner products:
\begin{eqnarray*}
(f,g) &\rightarrow& \langle f | g \rangle \equiv\int^{\pi}_{0} f(\theta) g(\theta) d\mu_{d}(\theta), \\
(f,g) &\rightarrow& \langle f | g \rangle_{\epsilon} \equiv \int^{\pi}_{\epsilon} f(\theta) g(\theta) d\mu_{d}(\theta),
\end{eqnarray*}
where $d\mu_{d}(\theta)$ are the measures in polar ($d=2$) and
spherical coordinates ($d=3$):
\begin{equation} 
\label{eq:mesure23}
d\mu_{2}(\theta) =  \frac{d\theta}{\pi} , \qquad   d\mu_{3}(\theta) = \frac{\sin \theta}{2}  d\theta.
\end{equation}
With these definitions, the eigenvectors $V_{n}(\theta)$ are orthonormal
\begin{equation} 
\label{eq:ortho}
\langle V_{n} | V_{m} \rangle =  \delta_{nm}. 
\end{equation}
We now use the boundary conditions (\ref{eq:bcrefl}-\ref{eq:bcadsorb})
to determine the coefficients $\{\alpha_{n}\}_{n}$ defining
$t_2(r,\theta)$ in Eq. (\ref{eq:t2_def}).

(i) The reflecting boundary condition (\ref{eq:bcrefl}) reads
\begin{eqnarray}
\beta_{0} \; \pa_{r} f_{0}(r)_{\lvert R_{c}} + \frac{1}{D_{2}} \pa_{r} \hat{f}(r)_{\lvert R_{c}} + \sum^{\infty}_{n=1}  
\left( \alpha_{n} \pa_{r} f_{n} + \alpha_{-n} \pa_{r} f_{-n} \right)_{\lvert R_{c}} V_{n} ( \theta ) = 0, 
\end{eqnarray}
which, using the orthogonality in Eq. (\ref{eq:ortho}), leads to the
following relations
\begin{eqnarray} 
\label{eq:alpha_n}  
\beta_{0} = -\frac{1}{D_{2}}\left(\frac{\pa_{r} \hat{f}(r)}{\pa_{r} f_{0}(r)}\right)_{\lvert r=R_{c}} , \qquad
\alpha_{n} \; \pa_{r} f_{n}(r)_{\lvert r=R_{c}}  &=& - \alpha_{-n} \; \pa_{r} f_{-n}(r)_{\lvert r=R_{c}}.
\end{eqnarray}
Note that, in the case $R_{c}=0$, the first condition reads
$\beta_{0}=0$. Indeed, if $\beta_{0}$ were non zero, the MFPT of a
molecule initially at the origin would diverge (by definition of the
function $f_{0}$).

(ii) Substituting Eq. (\ref{eq:t2_def}) into the radiative boundary
condition Eq. (\ref{eq:adsorption}), projecting it onto the basis
$V_n(\theta)$ and using Eq. (\ref{eq:alpha_n}), we obtain two
supplementary conditions:
\begin{eqnarray}
\label{eq:alpha0general}
\alpha_{0} - \frac{1}{D_{2}}\left(\frac{\pa_{r} \hat{f}(r)}{\pa_{r} f_{0}(r)}\right)_{\lvert r=R_{c}} 
\left(f_{0}(R) + \frac{1}{k} \pa_{r} f_{0}(R)\right) + \frac{1}{D_{2}} \left(\hat{f}(R) + \frac{1}{k} \pa_{r} \hat{f}(R)\right)
	&=& \langle t_{1} | 1 \rangle,   \\
\label{eq:alphangeneral}
\alpha_{n} \left[ f_{n}(R) + \frac{1}{k} \pa_{r} f_{n}(R) -  \left( \frac{\pa_{r} f_{n}(r)}{\pa_{r} f_{-n}(r)}\right)_{\lvert r=R_{c}}
 \left( f_{-n}(R)+ \frac{1}{k} \pa_{r} f_{-n}(R) \right) \right]
	&=&  \langle t_{1}| V_{n} \rangle  \quad (n>0) . 
\end{eqnarray}

On the other hand, the radiative boundary condition in
Eq. (\ref{eq:adsorption}) and the operator $L$ defined in
Eq. (\ref{def:l}) allow one to rewrite Eq. (\ref{eq:temps1}) as
\begin{equation*}  
 -\Delta_{\theta} t_{1}(\theta) = \frac{\omega^2}{\lambda} + \omega^2 (Lt_2)(R)  \qquad  (\epsilon < \theta < \pi),
\end{equation*}
which becomes, using Eqs. (\ref{eq:t2_def}, \ref{eq:alpha_n},
\ref{eq:alpha0general}, \ref{eq:alphangeneral}),
\begin{eqnarray}
\label{eq:diffgenerale}
- \Delta_{\theta} t_{1}(\theta)	&=& \omega^{2} T
+ \omega^{2} \sum^{\infty}_{n=1} X_{n} \langle t_{1} | V_{n} \rangle V_{n}(\theta)   \qquad  (\epsilon < \theta < \pi),
\end{eqnarray}
where
\begin{eqnarray} 
\label{eq:T} 
T	&\equiv& \frac{1}{\lambda}  + \frac{\eta_{d}}{D_{2}}, \\
\label{eq:eta}
\eta_{d}	&\equiv& -\left(\frac{\pa_{r} \hat{f}(r)}{\pa_{r} f_{0}(r)}\right)_{\lvert r=R_{c}} L f_{0}(R)
	    + L \hat{f}(R),  \\
\label{eq:Xn}
X_{n}	&\equiv& \frac{ L f_{n}(R) - \left(\frac{\pa_{r} f_{n}(r)}{\pa_{r} f_{-n}(r)}\right)_{\lvert r=R_{c}}
L f_{-n}(R)}{f_{n}(R) + \frac{1}{k} \pa_{r} f_{n}(R) - \left(\frac{\pa_{r} f_{n}(r)}{\pa_{r} f_{-n}(r)}\right)_{\lvert r=R_{c}}
 \left( f_{-n}(R) + \frac{1}{k} \pa_{r} f_{-n}(R) \right)} \quad (n \geq 1) \label{xcarac}.
\end{eqnarray}
In Appendix \ref{sec:etaMFTP} we identify the quantity $\eta_{d}/D_2$
as the mean first passage time on the sphere of radius $R$ for a
molecule initially at $r=R-a$.  Thus the time $T$ is the sum of a mean
exploration time $\eta_{d}/D_2$ and a mean ``exploitation'' time
$1/\lambda$.

(iii) The absorbing boundary condition (\ref{eq:bcadsorb}) and the
relation $t'_{1}(\pi)=0$ which comes from the invariance of $t_1$
under the symmetry $\theta \rightarrow 2\pi- \theta$, lead after
integration of Eq. (\ref{eq:diffgenerale}) to
\begin{equation} 
\label{eq:t1distribution}
t_{1}(\theta) = \begin{cases} \omega^2 T \; g_{\epsilon}(\theta) + \omega^2 \sum^{\infty}_{n=1} \frac{X_{n}}{\rho_{n}} 
\langle V_{n}|t_{1}\rangle _\epsilon\{V_{n}(\theta)-V_{n}(\epsilon)\} \quad (\epsilon < \theta < \pi) , \cr
0 \hskip 77mm  (0 \leq \theta \leq \epsilon), \end{cases}
\end{equation}
where $\rho_{n}$ is defined in Eq. (\ref{eq:rho}) and
$g_{\epsilon}(\theta)$ is the solution of the problem:
\begin{equation}
\label{eq:geps_def}
\Delta_{\theta} g_{\epsilon}(\theta) =-1, \quad {\rm with}\quad   g_{\epsilon}(\epsilon)=0\quad  
{\rm and}\quad  g'_{\epsilon}(\pi)=0.
\end{equation}
Note that $R^2 g_{\epsilon}(\theta)/D_1$ represents the MFPT to the
target when $\lambda=0$, i.e. in absence of desorption events, hence
$g_{\epsilon}(\theta)$ is well known:
\begin{equation}
g_{\epsilon}(\theta) = \begin{cases}   \displaystyle \frac{1}{2}(\theta-\epsilon)(2\pi-\epsilon-\theta)  \quad (d = 2), \cr
\displaystyle  \ln\left(\frac{1-\cos(\theta)}{1-\cos(\epsilon)}\right) \hskip 9.5mm (d = 3) . \end{cases}
\end{equation}
Equivalently, Eq.(\ref{eq:t1distribution}) reads 
\begin{equation}
\psi(\theta) = \begin{cases} \displaystyle g_{\epsilon}(\theta)
    + \omega^2 \sum^{\infty}_{n=1} \frac{X_{n}}{\rho_{n}} \langle V_{n}|\psi\rangle_\epsilon \{V_{n}(\theta)-V_{n}(\epsilon)\} 
\quad (\epsilon < \theta < \pi), \cr  0   \hskip 66mm  (0 \leq \theta \leq \epsilon), \end{cases}
\end{equation}
where $\psi(\theta) \equiv t_{1}(\theta)/(\omega^2 T)$ is a
dimensionless function.

\subsection{Exact solution}

The function $\psi(\theta)$ can be developed on the basis of functions
$\{V_{n}(\theta)-V_{n}(\epsilon)\}_{n}$,
\begin{equation*} 
\psi(\theta) = g_{\epsilon}(\theta) + \sum^{\infty}_{n=1} d_{n} \{V_{n}(\theta)-V_{n}(\epsilon)\}  \quad (\epsilon < \theta < \pi),
\end{equation*}
with coefficients $\{d_{n}\}_{n \geq 1}$ to be determined.  Due to
Eq. (\ref{eq:t1distribution}), the vector $\bm{d}= \{ d_{n} \}_{n\geq
1}$ is a solution of the equation
\begin{equation} 
\label{eq:relationdn} 
\sum^{\infty}_{n=1} d_{n} \{ V_ {n}(\theta)-V_ {n}(\epsilon) \} = 
\omega^2 \sum^{\infty}_{n=1} \left( U_{n} +  \sum^{\infty}_{m=1} Q_{n,m} d_{m} \right) \{ V_ {n}(\theta)-V_ {n}(\epsilon) \},
\end{equation} 
where we have defined the vectors $\bm{U}$ and $\bm{\xi}$ by their
$n$-th coordinates:
\begin{eqnarray}
\label{eq:defU}
U_{n} &\equiv& \frac{X_{n}}{\rho^{2}_{n}}~ \xi_{n} , \qquad  
\xi_{n} \equiv \rho_{n} ~ \langle g_{\epsilon}(\theta)|V_{n}(\theta)\rangle_{\epsilon}  \qquad (n \geq 1),
\end{eqnarray}
and the matrices $Q$ and $I_\epsilon$ by their elements:
%
\begin{eqnarray} 
\label{eq:defQ}
Q_{n,m} &\equiv& \frac{X_{n}}{\rho_{n}}~ I_{\epsilon}(n,m) , \qquad 
I_{\epsilon}(n,m)  \equiv \langle V_{n}(\theta)|V_{m}(\theta)-V_{m}(\epsilon)\rangle_{\epsilon} \qquad (m\geq 1,~ n\geq 1).
\end{eqnarray}
As Eq. (\ref{eq:relationdn}) is satisfied for all $\theta \in
(\epsilon,\pi)$, the coefficients $d_n$ can be found by inverting the
underlying matrix equation as
\begin{equation}
\label{eq:dn}
d_{n} = \left[\omega^2 \left(I - \omega^2 Q\right)^{-1} U\right]_{n}.
\end{equation}

The MFPT $t_1(\theta)$ can be explicitly rewritten as
\begin{equation} 
\label{eq:t1_final}
t_1(\theta) = \begin{cases}  \displaystyle \omega^2 T \left[g_{\epsilon}(\theta) + 
\sum^{\infty}_{n=1} d_{n} \{V_{n}(\theta)-V_{n}(\epsilon)\} \right] \quad (\epsilon < \theta < \pi), \cr
0 \hskip 60mm (0\leq \theta \leq \epsilon) . \end{cases}
\end{equation}
The averaged MFPT $\langle t_1 \rangle$ which is defined by averaging
over a uniform distribution of the starting point, is then easily
obtained as
%
\begin{equation}
\label{eq:searchtime} 
 \left\langle t_{1} \right\rangle \equiv \int^{\pi}_{0} t_{1}(\theta) d\mu_{d}(\theta)
=\omega^2T\left(\langle g_\epsilon|1\rangle_\epsilon+\sum^{\infty}_{n=1} d_{n} \xi_{n} \right),
\end{equation}
where we have used the following relation
\begin{equation*} 
\langle V_{n}(\theta)-V_{n}(\epsilon) | 1 \rangle_{\epsilon} 
=  - \langle V_{n}(\theta)-V_{n}(\epsilon) | \Delta_{\theta} g_{\epsilon}(\theta) \rangle_{\epsilon} = 
\rho_{n} \langle V_{n}(\theta)| g_{\epsilon}(\theta)\rangle_{\epsilon} = \xi_{n}.
\end{equation*}

Finally, the MFPT $t_2(r,\theta)$ is given by Eq. (\ref{eq:t2_def}),
in which the coefficients $\beta_0$ and $\alpha_n$ are obtained from
Eqs. (\ref{eq:alpha_n}, \ref{eq:alpha0general},
\ref{eq:alphangeneral}):
\begin{equation} 
\label{eq:t2_final}  
\begin{split}
t_{2}(r,\theta) & = \langle t_1 \rangle + \frac{\eta_d}{D_2} + \frac{1}{D_{2}} \biggl(\hat{f}(r) - \hat{f}(R-a)\biggr) 
- \frac{1}{D_2} \left(\frac{\partial_r \hat{f}}{\partial_r f_0}\right)_{r=R_c} \biggl(f_0(r) - f_0(R-a)\biggr) \\
& + \sum^{\infty}_{n=1} \alpha_{n} V_{n}(\theta)\biggl\{f_{n}(r)  - \left(\frac{\partial_r f_n}{\partial_r f_{-n}}\right)_{r=R_c}
f_{-n}(r) \biggr\}, \\
\end{split}
\end{equation} 
with
\begin{equation*}
\alpha_n = \frac{T \rho_n d_n}{L f_n(R) - \left(\frac{\partial_r f_n}{\partial_r f_{-n}}\right)_{r=R_c} L f_{-n}(R)}  \qquad (n \geq 1).
\end{equation*}
 
Table \ref{tab:Vtable} summarizes the quantities which are involved in
Eqs. (\ref{eq:t1_final}, \ref{eq:searchtime}, \ref{eq:t2_final}) and
independent of the detail of the radial bulk dynamics.  In turn, the
quantities $T$, $\eta_d$ and $X_n$ are expressed by Eqs. (\ref{eq:T},
\ref{eq:eta}, \ref{eq:Xn}) through the functions $\hat{f}$, $f_0$ and
$f_n$ and thus depend on the specific dynamics in the bulk phase and
will be discussed in Sec. \ref{sec:particular} for several particular
examples.

A numerical implementation of the exact solutions in
Eqs. (\ref{eq:t1_final}, \ref{eq:searchtime}, \ref{eq:t2_final})
requires a truncation of the infinite-dimensional matrix $Q$ to a
finite size $N\times N$.  After a direct numerical inversion of the
truncated matrix $(I-\omega^2 Q)$ in Eq. (\ref{eq:dn}), the MFPTs from
Eqs. (\ref{eq:t1_final}, \ref{eq:searchtime}, \ref{eq:t2_final}) are
approximated by truncated series (with $N$ terms).  We checked
numerically that the truncation errors decay very rapidly with $N$.
In a typical case of moderate $\omega < 100$, the results with $N =
100$ and $N = 200$ are barely distinguishable.  In turn, larger values
of $\omega$ (or $\lambda$) may require larger truncation sizes.  In
the following examples, we used $N = 200$.  In spite of the
truncation, we will refer to the results obtained by this numerical
procedure as {\it exact solutions}, as their accuracy can be
arbitrarily improved by increasing the truncation size $N$.  These
exact solutions will be confronted to approximate and perturbative
solutions described in the next subsections.

\begin{table}
 \begin{center}
 \begin{tabular}{|c|c|c|}
  \hline
Expressions & 2D & 3D \\
  \hline
$V_{n}(\theta)$ & $ \begin{cases} 1 \hskip 17mm (n=0)  \cr \sqrt{2} \cos(n \theta) \quad (n > 0) \end{cases}$ & $ \sqrt{2n+1} ~P_{n}(\cos \theta)$\\
  \hline
$\rho_{n}$ & $n^2$ & $n(n+1)$\\
  \hline
$d\mu_{d}(\theta)$ &  $d\theta/\pi$  & $\sin \theta \; d\theta/2$
\\
  \hline
$g_{\epsilon}(\theta)$ & $\frac{1}{2} (\theta-\epsilon)(2\pi-\epsilon-\theta)$ & $\ln\left(\frac{1-\cos(\theta)}{1-\cos(\epsilon)}\right)$\\
  \hline
$\langle g_{\epsilon}|1 \rangle_{\epsilon} $ & $\frac{1}{3 \pi} (\pi-\epsilon)^3$ 
& $\log\left(\frac{2}{1-\cos \epsilon}\right) - \frac{1+\cos \epsilon}{2}$\\
  \hline
$\langle g_{\epsilon}|V_{n} \rangle_{\epsilon} $ & 
{\small{$- \frac{\sqrt{2}}{ \pi n^{2}} \{ (\pi - \epsilon) \cos(n \epsilon) + \sin(n \epsilon)/n \}$}} & 
{\small{$- \frac{\sqrt{2n+1}}{2} \frac{1}{n(n+1)} \{ \left(1+ \frac{n \cos \epsilon}{n+1} \right) P_{n}(\cos \epsilon)
 + \frac{ P_{n-1}(\cos \epsilon)}{n+1} \}$}} \\
  \hline
$I_{\epsilon}(n,n), n\geq 1 $ & $ \frac{1}{\pi} \left(\pi - \epsilon + \frac{\sin 2 n \epsilon}{2 n} \right)$ & 
$\frac{2n+1}{2} \left( - P_n(u) \frac{u P_n(u) - P_{n-1}(u)}{n+1} + \frac{F_n(u) + 1}{2n+1} \right)$\\
  \hline
$I_{\epsilon}(n,m \neq n) $ & $\frac{2}{\pi} \frac{\cos(n\epsilon) \frac{\sin (m\epsilon)}{m} - \cos (m\epsilon) 
\frac{\sin (n\epsilon)}{n}}{n^2 - m^2} ~m^2$ & 
$\frac{\sqrt{2n+1}\sqrt{2m+1}}{2} m \frac{(n-m) u P_m(u) P_n(u) + (m+1)P_m(u) P_{n-1}(u) - (n+1)P_n(u)P_{m-1}(u)}{(n+1)[m(m+1)-n(n+1)]}$ \\
$m,n\geq 1$ & & where $u \equiv \cos\epsilon$ and function $F_n(u)$ is given in Appendix \ref{sec:ftable} \\
\hline
\end{tabular}
 \end{center}
\caption{
Summary of formulas for computing the vector $\bm{\xi}$ and the matrix
$Q$ in Eqs. (\ref{eq:defU}, \ref{eq:defQ}) that determine the
coefficients $d_n$ according to Eq. (\ref{eq:dn}).}
\label{tab:Vtable}
\end{table}

\subsection{Are bulk excursions beneficial?}

Before considering these perturbative and approximate solutions, we address the  important issue of determining whether bulk excursions are
beneficial for the search.  This question can be answered by studying
the sign of the derivative of $\langle t_{1}\rangle$ with respect to
$\lambda$ at $\lambda=0$.  In terms of $\tilde{Q}=-Q R^{2}/D_{1}$, the
MFPT from Eq.  (\ref{eq:searchtime}) reads
\begin{equation}
\label{eq:t1matrix} 
\langle t_{1} \rangle = \frac{R^4}{D^{2}_{1}}
(1+\lambda \eta_{d}/D_{2}) \left[\frac{D_{1}}{R^2} \langle
g_{\epsilon}|1\rangle_{\epsilon} + \biggl(\bm{\xi} \cdotp \lambda (I+\lambda
\tilde{Q})^{-1}\bm{U}\biggr) \right].  
\end{equation} 
The derivative of $\langle t_{1}\rangle$ with respect to $\lambda$ is
\begin{equation} 
\label{t1deriveparlambda}
\pd{\langle t_{1} \rangle}{\lambda} = \frac{R^4 \eta_{d}}{D^{2}_{1}} \left[ \frac{D_{1}}{D_{2} R^2} \langle g_{\epsilon}|1\rangle_{\epsilon} +  
\left( \bm{\xi} \cdotp \frac{(\eta^{-1}_{d}+2\lambda)I + \lambda^{2}\tilde{Q}}{(I+\lambda \tilde{Q})^{2}} \bm{U} \right)\right] .
\end{equation}
If the derivative is negative at $\lambda =0 $, i.e.
\begin{equation}
\frac{D_{1}}{D_{2}} \leq -\frac{R^2}{\langle g_{\epsilon}|1\rangle_{\epsilon}} ~\frac{(\bm{\xi}\cdotp \bm{U})}{\eta_{d}},
\end{equation}
bulk excursions are beneficial for the search.  Explicitly, the
critical ratio of the bulk-to-surface diffusion coefficients, below
which bulk excursions are beneficial, is
\begin{equation} 
\label{eq:conditionD1D2generale}
\frac{D_{2c}}{D_{1}} = -\frac{\eta_d  \langle g_{\epsilon}|1 \rangle_{\epsilon}}{R^2} 
\left[\sum^{\infty}_{n=1} X_{n} \langle g_{\epsilon}|V_{n} \rangle^{2}_{\epsilon} \right]^{-1}.
\end{equation}

\subsection{Perturbative solution (small $\epsilon$ expansion)}

While Eq. (\ref{eq:t1_final}) for $t_{1}$ is exact, it is not fully
explicit since it requires either the inversion of the
(infinite-dimensional) matrix $I - \omega^2 Q$, or the calculation of
all the powers of $Q$. In this section, we give the first terms of  a small $\epsilon$ expansion of the MFPT, while in the next one we provide an approximate solution that improves in practice the range of validity of this perturbative solution. Both solutions rely 
on the orthogonality of functions $V_n$ in
the small target size limit $\epsilon\to 0$, which implies that the
matrix $Q$ is diagonal in this limit.  

More precisely, as
$V_{n}(\theta)=V_{n}(-\theta)$, necessarily $\pa_{\theta} V_{n}(0)=0$,
so that for $\epsilon$ close to zero and for all $\theta
\in [0,\epsilon]$, one has: $V_{n}(\theta) = V_{n}(0) +
O(\theta^{2})$.  As a consequence, the function $I_{\epsilon}(n,m)$
introduced in Eq. (\ref{eq:defQ}), reads for all $m,n\geq 1$ (see also
Appendix \ref{sec:ftable})
\begin{equation}
\label{eq:Ieps} 
I_{\epsilon}(n,m) \equiv \langle
V_{n}(\theta)|V_{m}(\theta)-V_{m}(\epsilon)\rangle_{\epsilon} =
\langle V_{n}(\theta)|V_{m}(\theta)\rangle-V_m(\epsilon)\langle
V_{n}(\theta)|1\rangle+O(\epsilon^3) =\delta_{nm}+O(\epsilon^3).
\end{equation}

The first terms of a small $\epsilon$ expansion of the MFPT can then
be exactly calculated.  Relying on the expansion Eq. (\ref{eq:Ieps}),
one can replace $I_\epsilon(n,n)$ by $1$
to get
in 2D
\begin{equation}
\label{eq:perturb2D} 
\frac{\langle t_{1}\rangle}{\omega^2T} = \left(\frac{\pi^2}{3}+2\omega^2\sum_{n=1}^\infty
\frac{X_n}{n^2(n^2 - \omega^2 X_n)}\right) - \pi \epsilon +
\left(1 - 2\omega^2\sum_{n=1}^\infty\frac{X_n}{n^2 - \omega^2 X_n}\right)\epsilon^2+O(\epsilon^3),
\end{equation}
and in 3D
\begin{equation}
\label{eq:perturb3D}
\frac{\langle t_{1}\rangle}{\omega^2T} = -2\ln(\epsilon/2)- \left(1 + \omega^2\sum_{n=1}^\infty
\frac{(2n+1) X_n }{n(n+1)(\omega^2 X_n - n(n+1))}\right) + O(\epsilon^2).
\end{equation}
The comparison of the perturbative solutions to the exact and
approximate ones is presented in Figs. \ref{fig:case1_k1},
\ref{fig:case1_lambda}, \ref{fig:case2}, \ref{fig:case3} and it is discussed
below.

\subsection{Approximate solution}

As mentioned above, we now provide an approximate solution that improves in practice the range of validity of the perturbative solution. 
This approximation relies on the fact that, due to Eq. (\ref{eq:Ieps}), the matrix $Q$ defined in Eq. (\ref{eq:defQ}) reads
\begin{equation}
\label{eq:QQ}
Q_{nm}= \delta_{mn} Q_{nn}+ O(\epsilon^{3}).
\end{equation}

\begin{figure}
\begin{center}
\includegraphics[width=80mm]{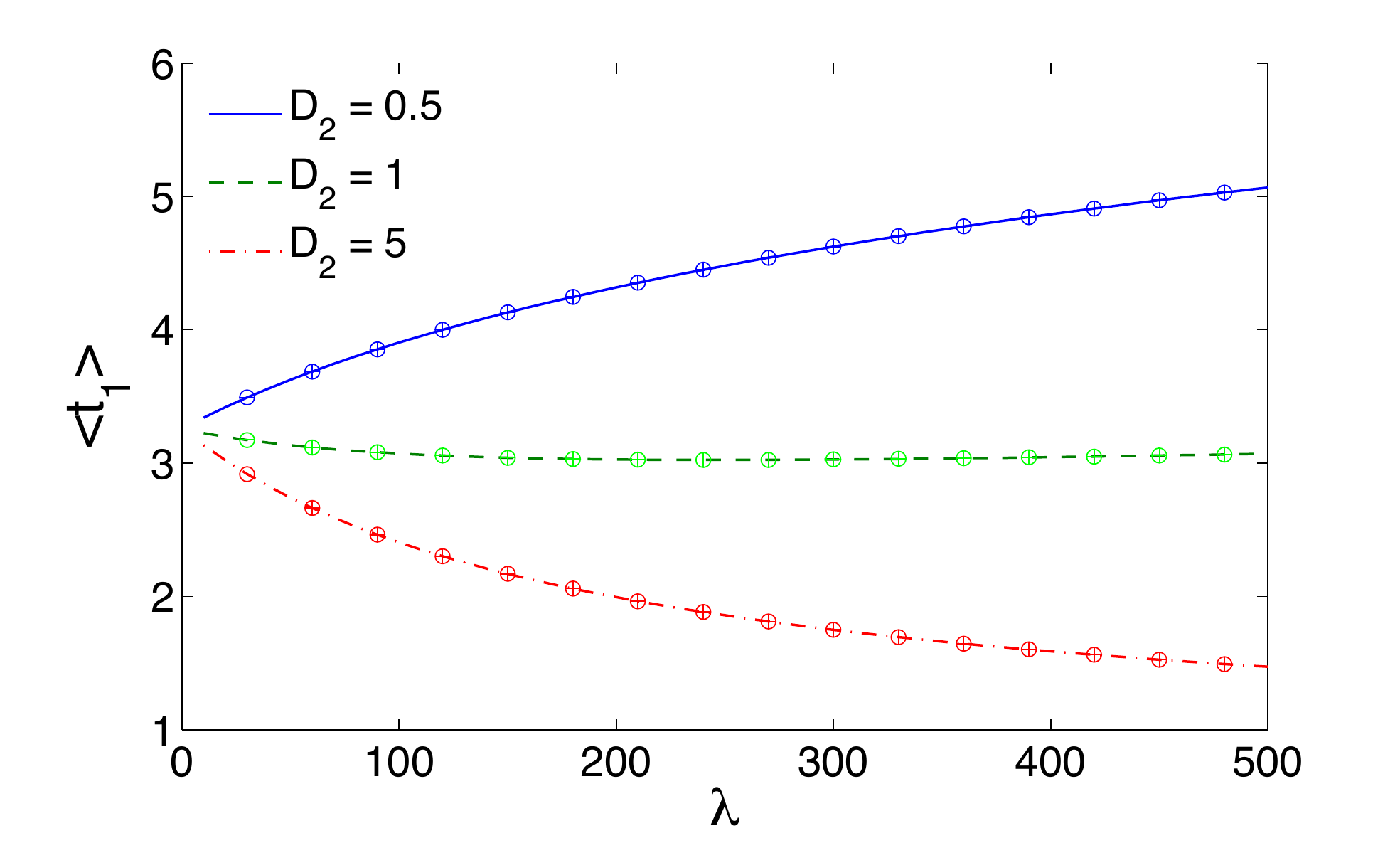}   \includegraphics[width=80mm]{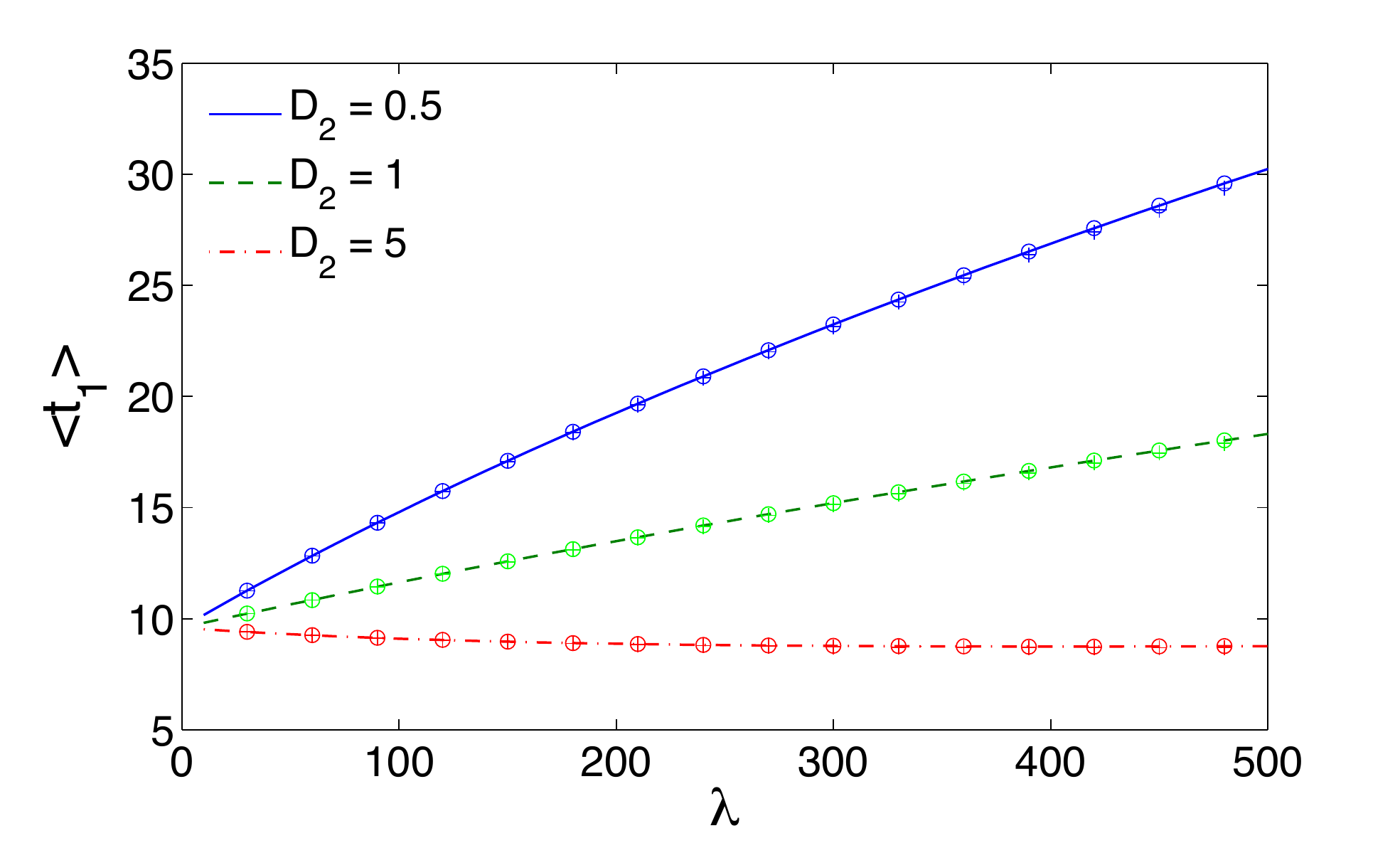}
\includegraphics[width=80mm]{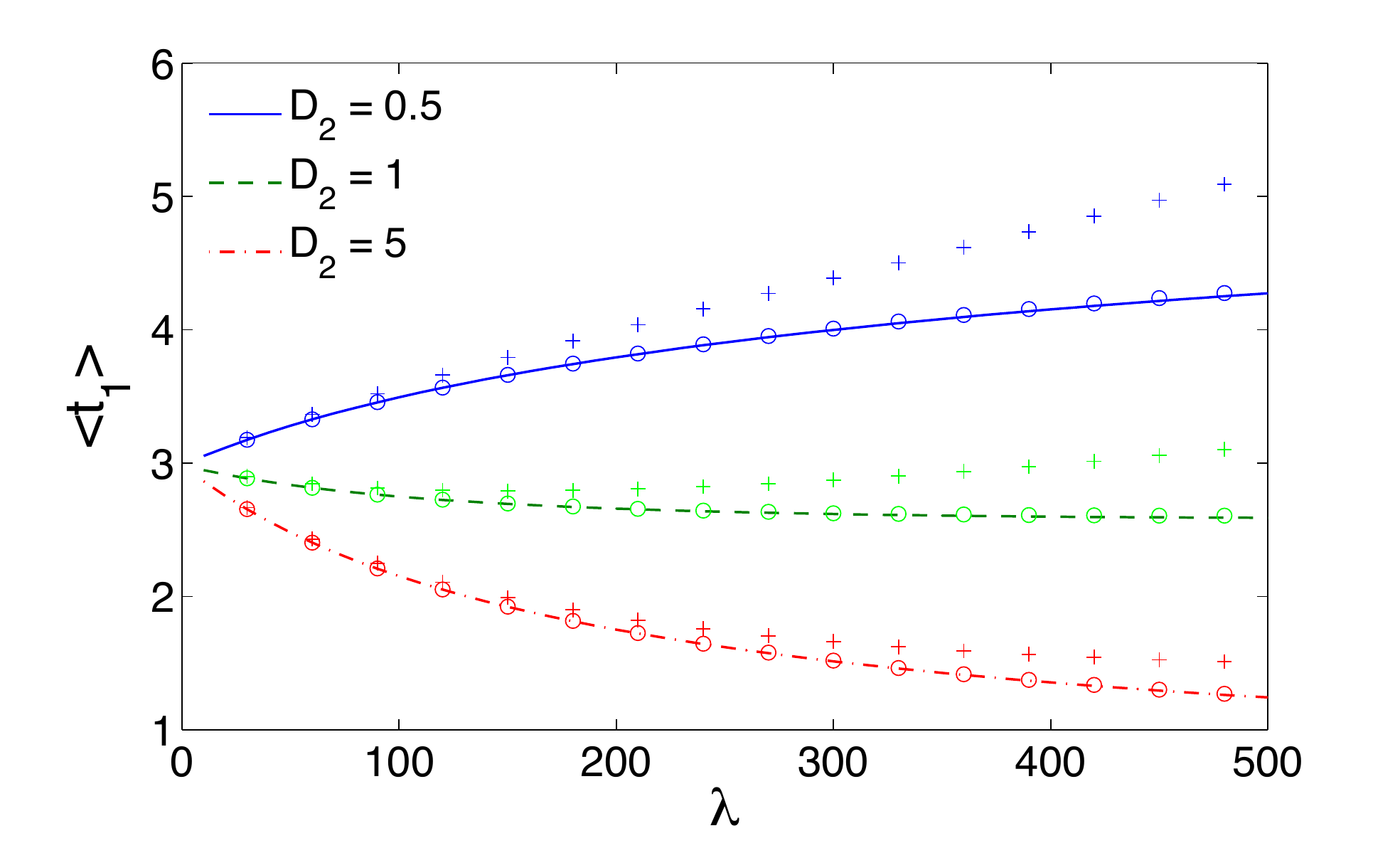}    \includegraphics[width=80mm]{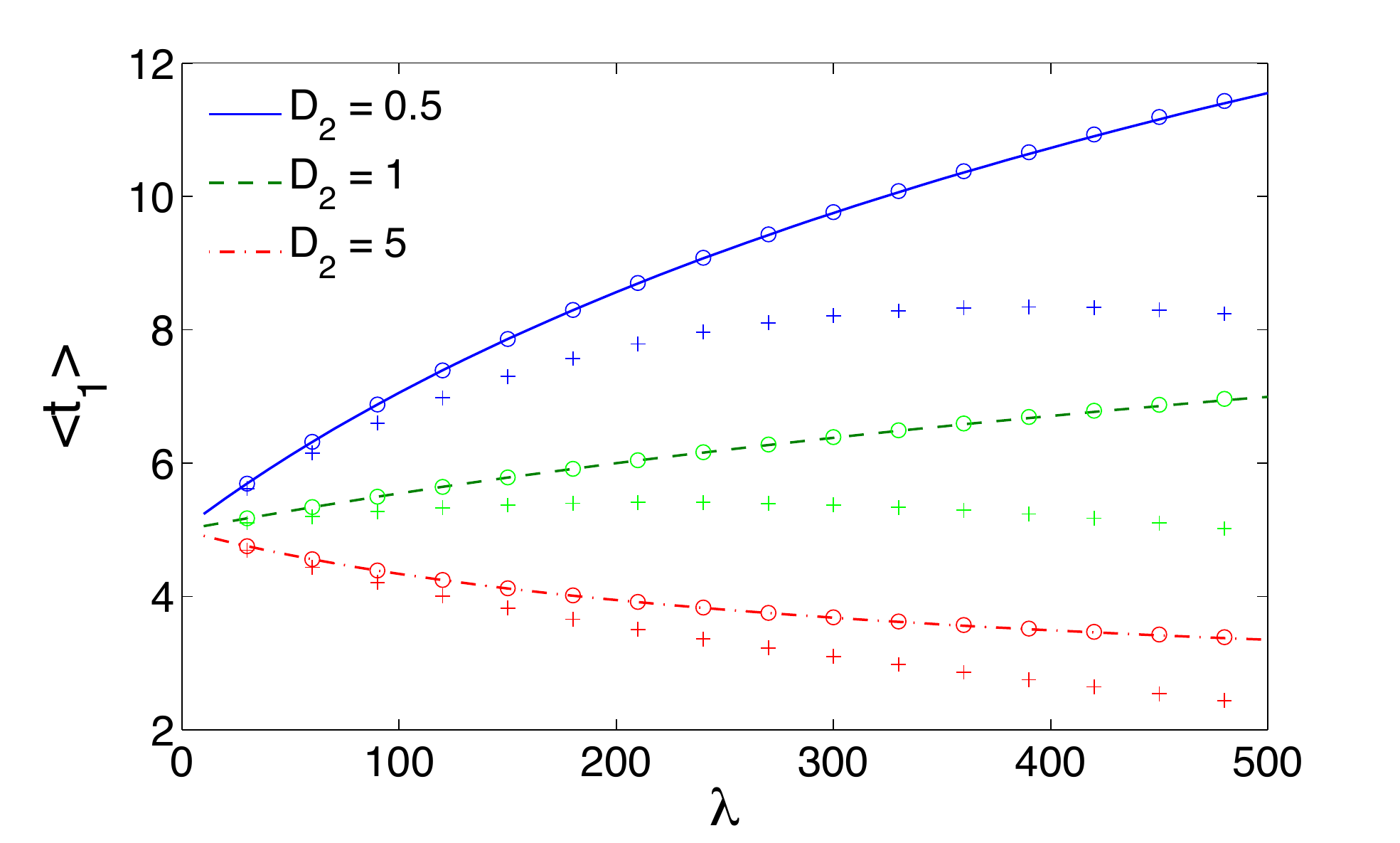}
\end{center}
\caption{
MFPT $\langle t_1\rangle$ as a function of the desorption rate
$\lambda$ for an annulus with the inner radius $R_c = 0.5$ and the
outer radius $R = 1$: comparison between the exact solution (lines),
approximate solution (circles) and perturbative solution (pluses) for
2D (left) and 3D (right), with $\epsilon = 0.01$ (top) and $\epsilon =
0.1$ (bottom).  The other parameters are: $D_1 = 1$, $a = 0.01$, $k =
\infty$, no bias ($V = 0$), and $D_2$ takes three values $0.5$, $1$
and $5$ (the truncation size is $N = 200$). }
\label{fig:case2}   
\end{figure}

Keeping only the leading term of this expansion, one gets
\begin{equation}
\label{eq:diag} 
d_{n} \approx  \frac{\omega^2 U_n}{1 - \omega^2 Q_{nn}} \quad (n\geq 1).
\end{equation}
From Eqs. (\ref{eq:defU}, \ref{eq:searchtime}, \ref{eq:diag}) we then
obtain the following approximation for the search time:
\begin{equation} 
\label{eq:t1general}
\langle t_{1}\rangle \approx \omega^2 T\left[ \langle g_{\epsilon}|1\rangle_{\epsilon} 
+ \omega^2 \sum^{\infty}_{n=1} 
\frac{X_{n} ~ \langle g_{\epsilon}|V_{n}\rangle^{2}_{\epsilon} }{1-\omega^2 \frac{X_{n}}{\rho_{n}  } I_{\epsilon}(n,n)} \right].
\end{equation}
Note that this expression is fully explicit as soon as the functions
$\hat{f}$, $f_0$ and $f_n$ defined in Eqs (\ref{eq:fhat}, \ref{eq:f0},
\ref{eq:fn_def}) are determined. In Section \ref{sec:particular}, we
will consider particular examples and write these functions
explicitly.  As we will show numerically, this approximation of $t_1$,
which was derived for small $\epsilon$, is in an excellent
quantitative agreement with the exact expression for a wide range of
parameters and even for large targets (see Figs. \ref{fig:case1_k1},
\ref{fig:case1_lambda}, \ref{fig:case2}, \ref{fig:case3}).

\section{Particular cases} 
\label{sec:particular}

We now show how the above theoretical approach can be applied to
various important examples.  The only quantities needed to obtain
fully explicit expressions of Eqs. (\ref{eq:searchtime},
\ref{eq:t1general}, \ref{eq:perturb2D}, \ref{eq:perturb3D}) are the
functions $\hat{f}$, $f_0$ and $f_n$ defined in Eqs. (\ref{eq:fhat},
\ref{eq:f0}, \ref{eq:fn_def}) which are involved in the definitions of
the quantities $T$ and $X_n$ according to Eqs. (\ref{eq:T},
\ref{eq:Xn}).  These quantities are listed in Table \ref{tab:ftable}
for the representative cases discussed in this section.  Throughout in
this section, all the quantities ($R$, $R_c$, $a$, $\epsilon$,
$\lambda$, $D_1$, $D_2$, $k$, $\langle t_1\rangle$) are written in
dimensionless units.  The physical units can be easily retrieved from
the definitions of these quantities.

\begin{table}
\begin{center}
\begin{tabular}{|c|c|c|c|}  \hline
Case & Quantity & 2D & 3D \\ \hline
No bias ($V = 0$)	 	& $\hat{f}$ 		& $-r^2/4$ 	& $-r^2/6$ 	\\
				& $f_{0}$		& $\ln r$	& $R/r$		\\
				& $f_{n}$		& $r^{n}$	& $r^{n}$ 	\\	
				& $f_{-n}$		& $r^{-n}$	& $r^{-n-1}$	\\
\hline
Velocity field: 		& $\hat{f}$ 		& $-r^2/(2 (2-\mu))$ 	& $-r^2/(2 (3-\mu))$ 	\\
$\vec{v}(r) = -\frac{\mu D_{2}}{r^2} \; \vec{r}$
				& $f_{0}$		& $[(r/R)^\mu-1]/\mu$	& $(r/R)^{\mu-1}/(1-\mu)$ 	\\
				& $f_{n}$		& $r^{\mu/2 + \gamma_n}$ & $r^{(\mu-1)/2 + \gamma_n}$ 	\\	
				& $f_{-n}$		& $r^{\mu/2 - \gamma_n}$ & $r^{(\mu-1)/2 - \gamma_n}$ 	\\
				&			& ($\gamma_n\equiv\sqrt{n^2 + \mu^2/4}$) & ($\gamma_n\equiv\sqrt{n(n+1) + (\mu-1)^2/4}$) \\
\hline
Sector of angle $\phi$		& $\hat{f}$ 		& $-r^2/4$ 		& $-r^2/6$  	\\
(no bias, $V = 0$)		& $f_{0}$		& $\ln r$ 	& $R/r$			\\
				& $f_{n}$		& $r^{n \pi/\phi}$ 	& $r^{-(1/2) + \gamma_n}$  \\
				& $f_{-n}$		& $r^{-n \pi/\phi}$ 	& $r^{-(1/2) - \gamma_n}$  \\
				&			&			& 
{\small{($\gamma_n \equiv \sqrt{n (n + 1) (\pi/\phi)^2 + 1/4}$)}} \\	
\hline
\end{tabular}
\end{center}
\caption{
Functions $\hat{f}$, $f_0$ and $f_n$ for several particular cases in
2D and 3D (see also Appendix \ref{sec:Vr2}). }
\label{tab:ftable}
\end{table}

\subsection{Zero bias ($V = 0$)}

\subsubsection{Exit problem for a perfect adsorption} \label{sec:perfectadsorption}

In the case of the exit problem with $R_{c}=0$, perfect adsorption ($k
= \infty$) and no bias, the formula (\ref{eq:t1general}) reproduces
the results of \cite{Benichou:2011a}.  The coefficient
$\eta_{d}/D_{2}$ is the mean first passage time to the sphere of
radius $R$, starting from $r=R-a$,
\begin{equation} \label{eq:etasimple}
\frac{\eta_{d}}{D_{2}} = \frac{a (2R-a)}{2 d}.
\end{equation}
From the expressions for the quantities $T$ and $X_{n}$,
\begin{equation*}
T = \frac{1}{\lambda} + \frac{R^2}{2 d D_{2}}(1-x^2) , \qquad   X_{n} = x^{n}-1,
\end{equation*}
we retrieve the approximate expressions for the MFPT in 2D
\begin{equation*} 
 \langle t_{1}\rangle \approx \frac{\omega^2 T}{\pi} \left[ \frac{1}{3} (\pi-\epsilon)^3
+ \frac{2 \omega^2}{\pi}  \sum^{\infty}_{n=1} \frac{x^{n}-1}{n^4} 
\frac{\left((\pi - \epsilon) \cos(n \epsilon) + \sin(n \epsilon)/n\right)^2}{1 - \frac{\omega^2}{\pi} 
\frac{x^{n}-1}{n^{2}} \left(\pi - \epsilon + \frac{\sin 2 n \epsilon}{2 n} \right)} \right],
\end{equation*}
and in 3D:
\begin{equation*} 
\langle t_{1}\rangle \approx \omega^2 T \left[ \ln\left(\frac{2}{1-\cos \epsilon}\right) - \frac{1+\cos \epsilon}{2}
+ \frac{\omega^2}{4} \sum^{\infty}_{n=1} \frac{(x^{n}-1)(2n+1)}{n^2(n+1)^2} 
\frac{\left( \left(1+ \frac{n \cos \epsilon}{n+1} \right) P_{n}(\cos \epsilon) + 
\frac{ P_{n-1}(\cos \epsilon)}{n+1}\right)^{2}}{1-\frac{\omega^2}{2}\frac{(x^{n}-1)(2n+1)}{n(n+1)} I_{\epsilon}(n,n)} \right].
\end{equation*}

We emphasize that bulk excursions can be beneficial for the MFPT even
for the bulk diffusion coefficient $D_2$ smaller than the surface
diffusion coefficient $D_1$ \cite{Benichou:2011a}.  This can be
understood qualitatively by the fact that bulk diffusion induces
flights towards remote and unvisited regions of the sphere $r=R$.  These
long-range hops can diminish the time for target encounter
(provided that the time spent in the bulk phase is not too large).

\subsubsection{Exit time for a partial adsorption}

We now give an explicit expression of the results (\ref{eq:t1general})
and (\ref{eq:conditionD1D2generale}) for a 2D exit problem with
$R_{c}=0$ and with an imperfect adsorption on the sphere of radius
$R$.  Using the expressions from Table \ref{tab:ftable}, the
coefficients $\eta_d$ and $X_{n}$ are
\begin{eqnarray}
\eta_d = \frac{R^2}{2 d}\left(1-x^2+ \frac{2}{kR}\right) , \qquad 
X_{n} = \frac{x^{n}-1-\frac{n}{kR}}{1 + \frac{n}{kR}}.
\end{eqnarray}
Thus the approximate MFPT in 2D reads 
\begin{equation} 
\label{t1d2:k}
\langle t_{1}\rangle \approx \frac{\omega^2 T}{\pi} \left[ \frac{1}{3} (\pi-\epsilon)^3
- \frac{2 \omega^2}{\pi}  \sum^{\infty}_{n=1} \frac{1-x^{n}+\frac{n}{kR}}{n^4 \left(1 + \frac{n}{kR}\right)} 
\frac{\left((\pi - \epsilon) \cos(n \epsilon) + \sin(n \epsilon)/n\right)^2}{1 - \frac{\omega^2}{\pi} 
\frac{x^{n}-1-\frac{n}{kR}}{n^{2} \left(1 + \frac{n}{kR}\right)} \left(\pi - \epsilon +\frac{\sin 2 n \epsilon}{2 n} \right)} \right],
\end{equation}
and the critical ratio of the bulk-to-surface diffusion coefficients
in Eq. (\ref{eq:conditionD1D2generale}), below which bulk excursions
are beneficial, takes the form
\begin{equation}
\frac{D_{2c}}{D_{1}} = \left(1-x^2+ \frac{2}{kR}\right) \frac{\pi (\pi-\epsilon)^3}{24} \left[\sum^{\infty}_{n=1}
 \frac{1-x^{n}+\frac{n}{kR}}{n^4 \left(1 + \frac{n}{kR}\right)} \left( (\pi - \epsilon) \cos(n \epsilon) + \sin(n \epsilon)/n \right)^2 \right]^{-1}.
\end{equation}
Similarly, one can write explicit formulas in 3D.

\begin{figure}
\begin{center}
\includegraphics[width=80mm]{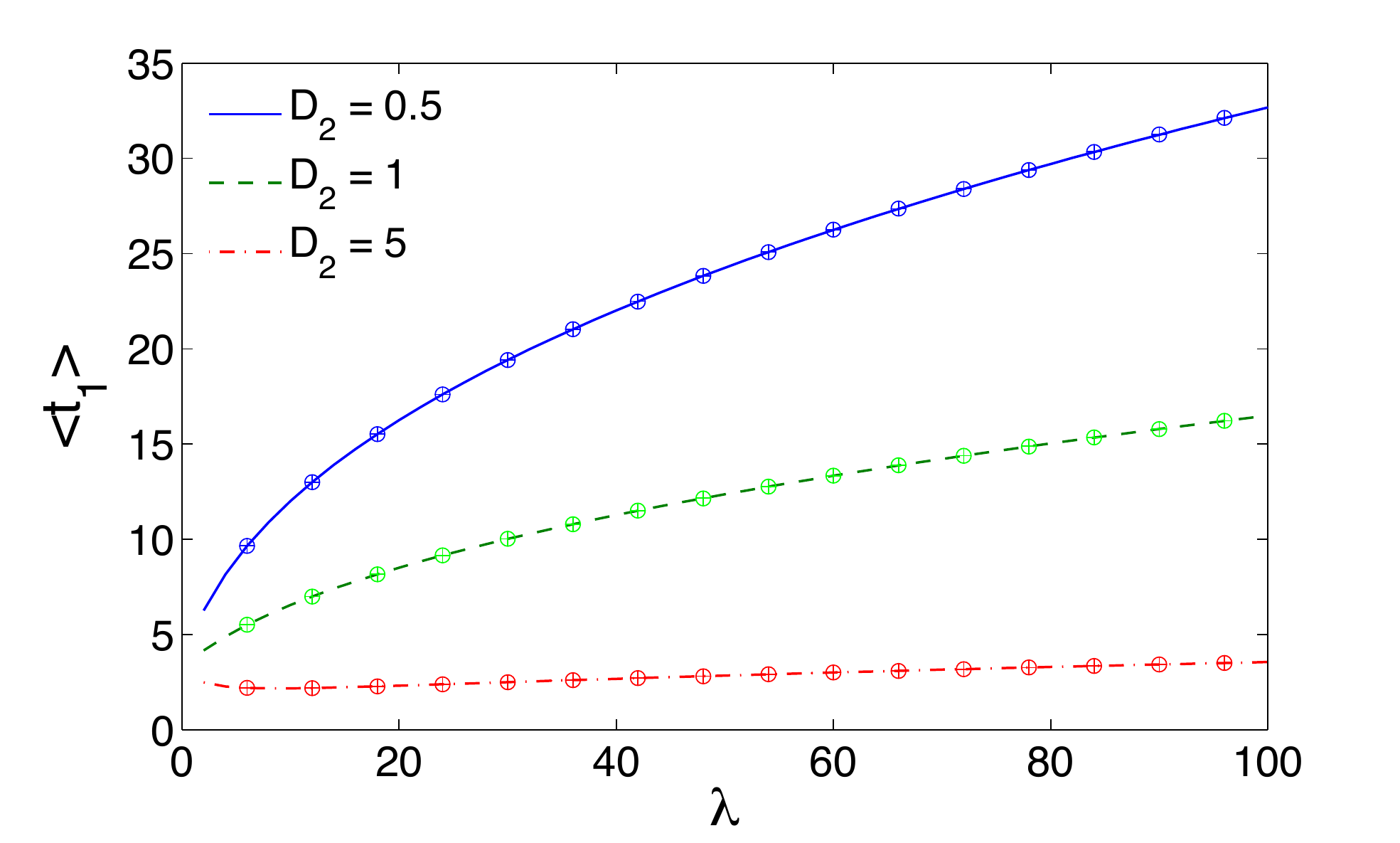}   \includegraphics[width=80mm]{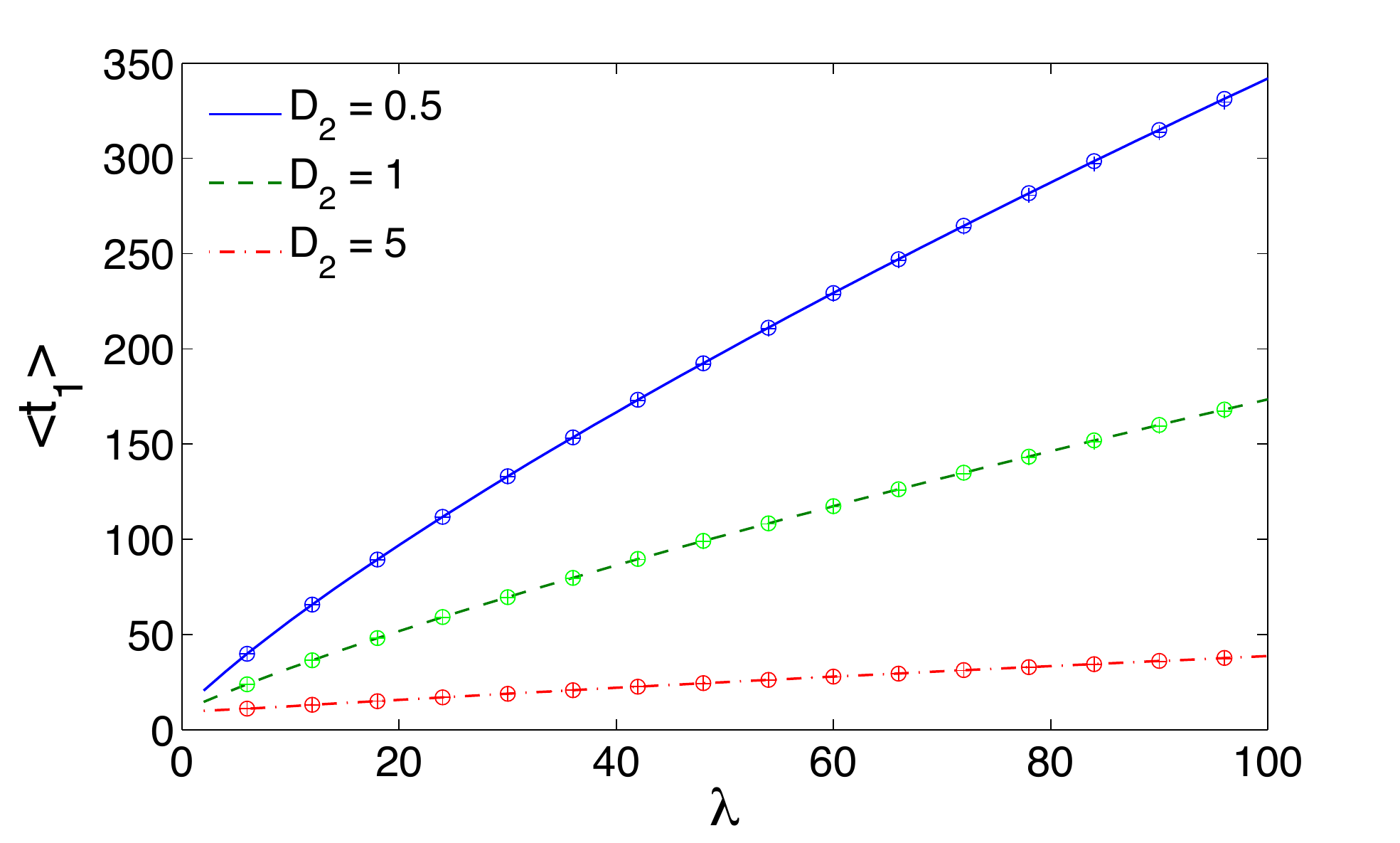}
\includegraphics[width=80mm]{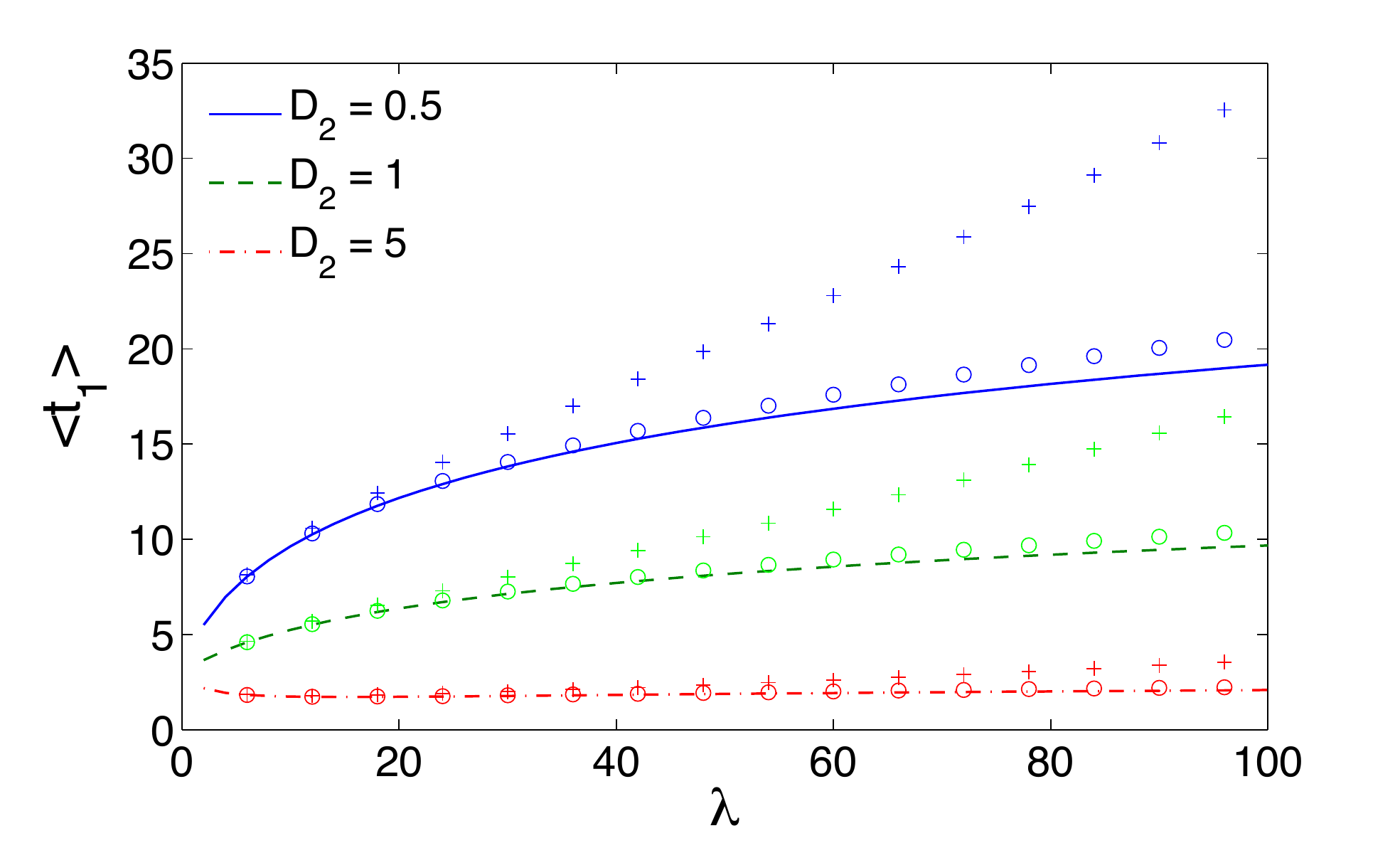}    \includegraphics[width=80mm]{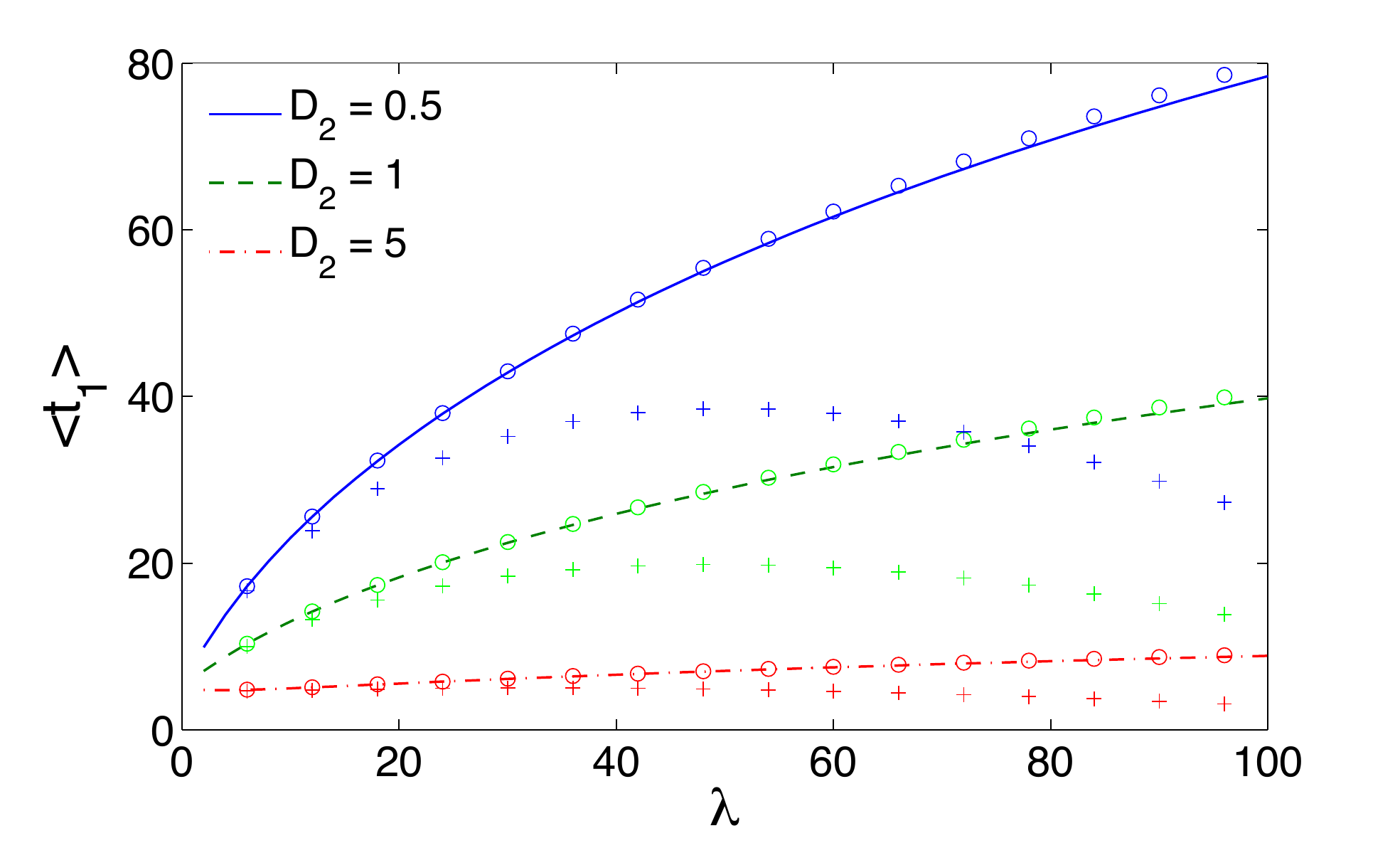}
\end{center}
\caption{
MFPT $\langle t_1\rangle$ as a function of the desorption rate
$\lambda$ for domains with partial adsorption $k = 1$: comparison
between the exact solution (lines), approximate solution (circles) and
perturbative solution (pluses) for 2D (left) and 3D (right), with
$\epsilon = 0.01$ (top) and $\epsilon = 0.1$ (bottom).  The other
parameters are: $R = 1$, $D_1 = 1$, $a = 0.01$, no bias ($V = 0$), and
$D_2$ takes three values $0.5$, $1$ and $5$ (the truncation size is $N
= 200$). }
\label{fig:case1_k1}   
\end{figure}

The MFPT as a function of the desorption rate $\lambda$ is shown on
Fig. \ref{fig:case1_k1} for different values of the bulk diffusion
coefficient $D_2$ and the target sizes $\epsilon = 0.01$ and $\epsilon
= 0.1$, both in two and three dimensions.  One can see that the
approximate solution (\ref{t1d2:k}) (shown by circles) accurately
follows the exact solution (shown by lines) for a wide range of
parameters.  In turn, the perturbative solutions in
Eqs. (\ref{eq:perturb2D}, \ref{eq:perturb3D}) (shown by pluses) are
accurate for small $\epsilon = 0.01$ but they deviate from the exact
ones for larger $\epsilon = 0.1$.

\begin{figure}
\begin{center}
\includegraphics[width=80mm]{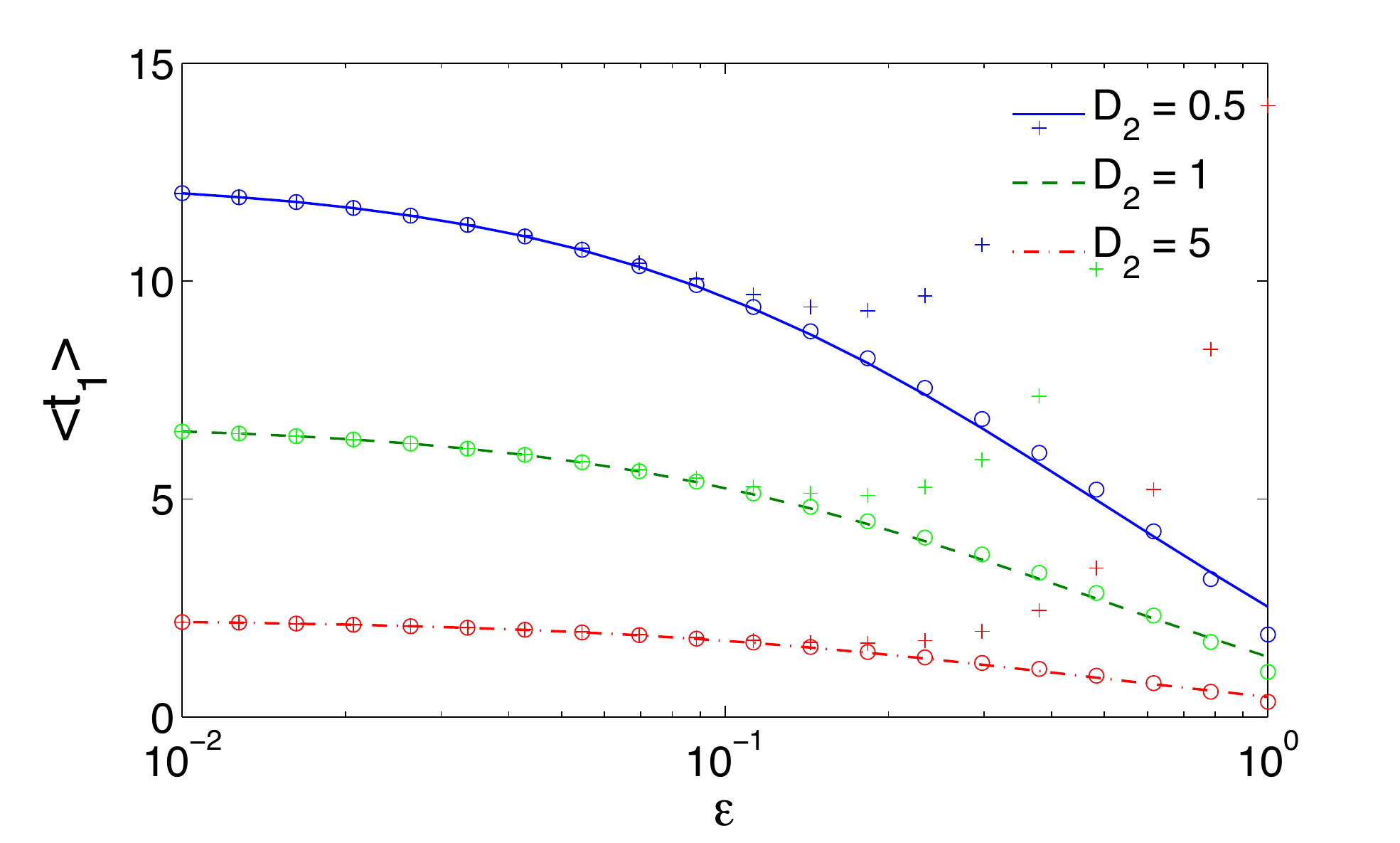}   \includegraphics[width=80mm]{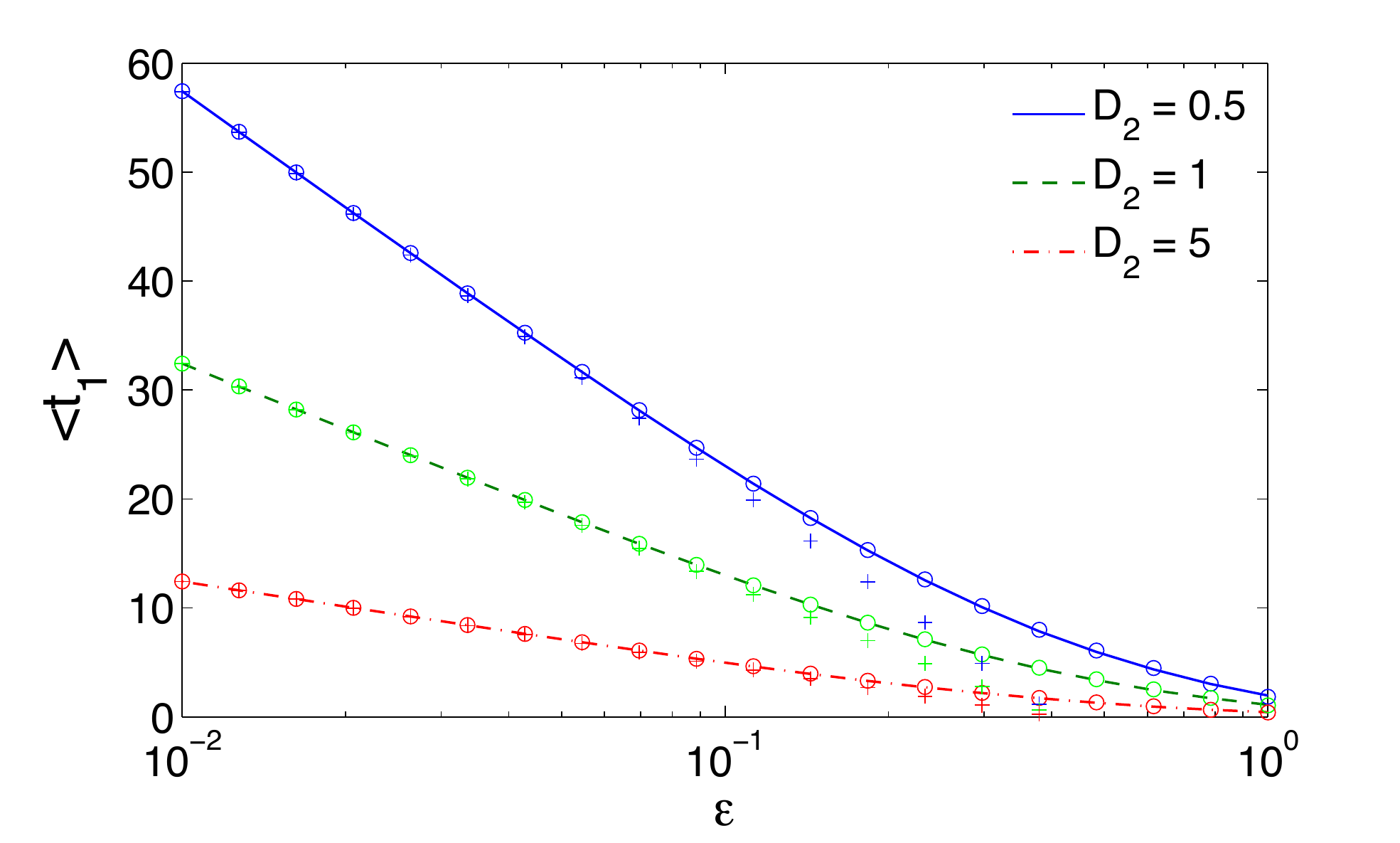}
\end{center}
\caption{
MFPT $\langle t_1\rangle$ as a function of the target size $\epsilon$
for domains with partial adsorption $k = 1$: Comparison between the
exact solution (lines), approximate solution (circles) and
perturbative solution (pluses) for 2D (left) and 3D (right).  The
other parameters are: $R = 1$, $D_1 = 1$, $a = 0.01$, $\lambda = 10$,
no bias ($V = 0$), and $D_2$ takes three values $0.5$, $1$ and $5$
(the truncation size is $N = 200$).  }
\label{fig:case1_lambda}   
\end{figure}

The quality of the approximate and perturbative solutions can also be
analyzed on Fig. \ref{fig:case1_lambda} which shows the MFPT as a
function of the target size $\epsilon$ (with a moderate value $\lambda
= 10$).  Once again, the approximate solution is very accurate for the
whole range of $\epsilon$, with a notable deviation only at $\epsilon$
close to $1$.  The perturbative solution starts to deviate for
$\epsilon \geq 0.1$ (as the desorption rate $\lambda$ appears in the
coefficients of the perturbative series, the validity range would of
course depend on $\lambda$ used).

\begin{figure}[h!]
	\centering
\includegraphics[width=0.45\textwidth]{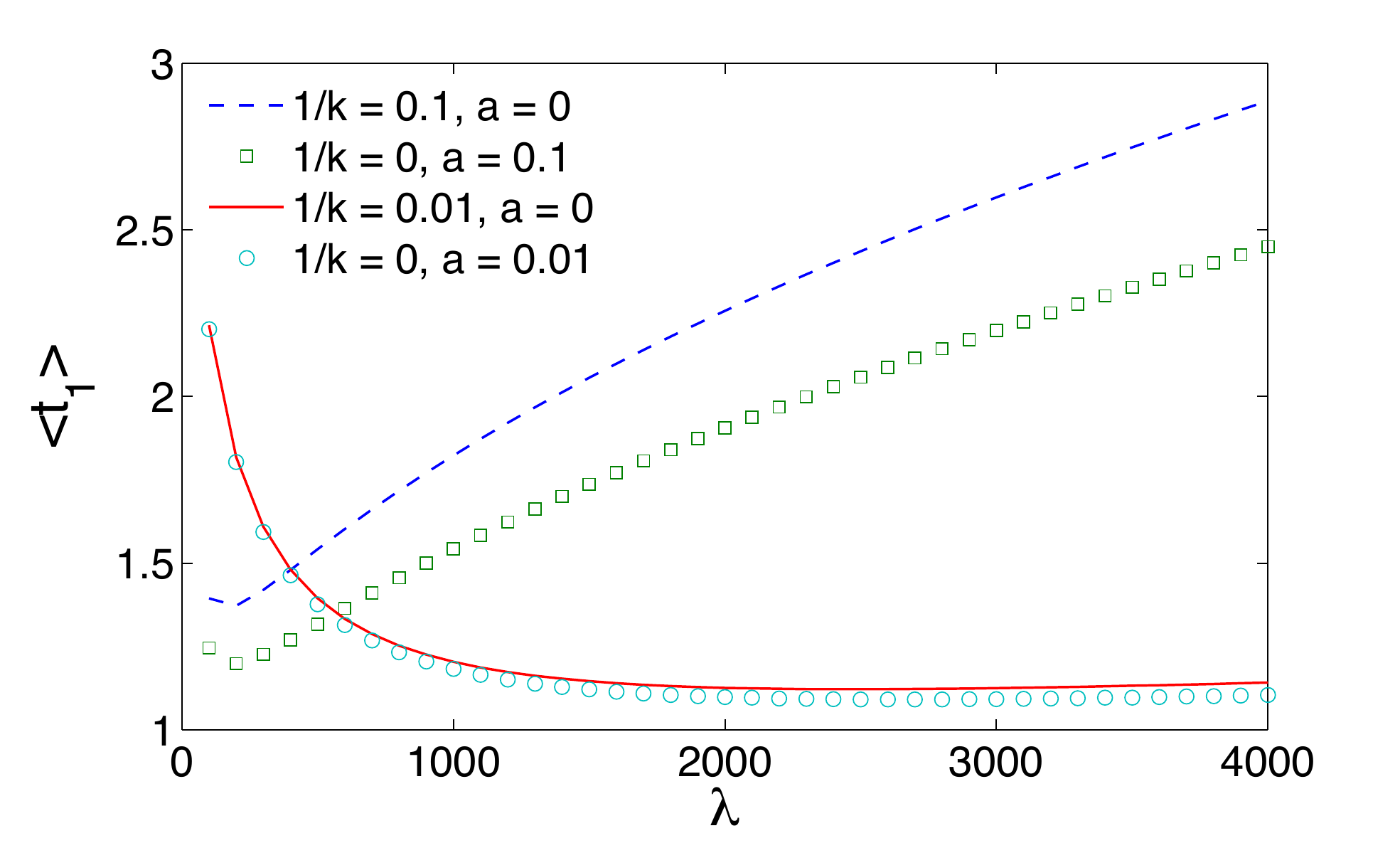}
\includegraphics[width=0.45\textwidth]{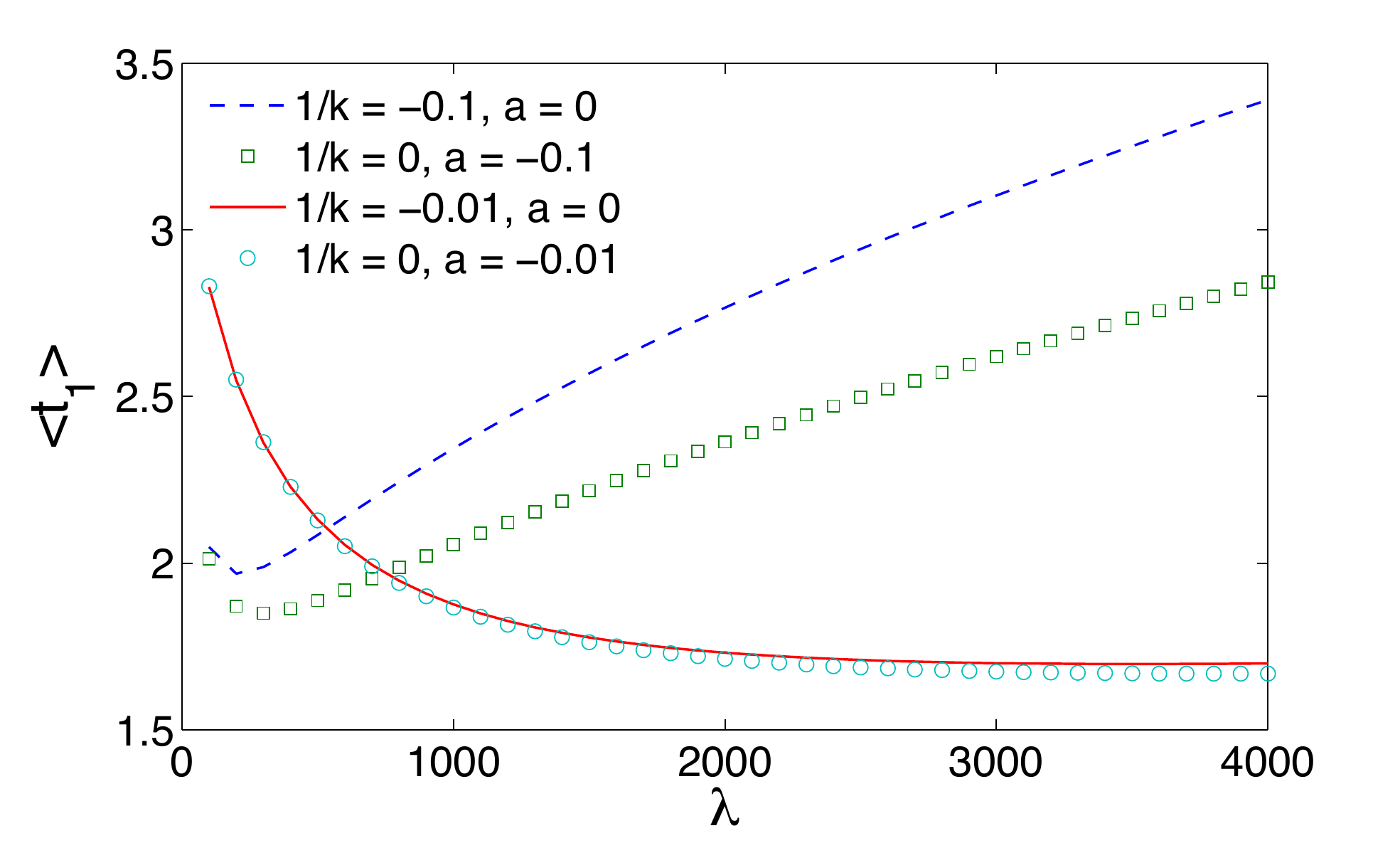}
\caption{ \label{fig:imperfectadsorption}
MFPT $\langle t_{1} \rangle$ computed through
Eq. (\ref{eq:searchtime}) as a function of the desorption rate
$\lambda$ for several combinations of the parameters $k$ and $a$, with
$R_{c}=0$ (left) and $R_{c}=\sqrt{2} > R = 1 $ (right).  The relation
between $1/k$ and $a$ is asymptotically valid for small $a$.  If both
values of $a$ and $1/k$ are close to zero, the MFPT (green line) tends
to be a constant which is equal to the MFPT on a segment of length
$2\pi$.  Here $d = 2$, $R = 1$, $D_1 = 1$, $D_2 = 5$, $\epsilon = 0$,
no bias ($V=0$) (the truncation size is $N=200$). }
\label{fig:t1vslambda(k)Rapport}   
\end{figure}

The situation of quasi-perfect adsorption ($kR\gg1$) can be shown to
be asymptotically equivalent with the case of short ejection distance
($a/R\ll1$), as illustrated on Fig. \ref{fig:imperfectadsorption}.

\subsubsection{Reflecting boundary and entrance time}

Now we provide an explicit form for Eqs. (\ref{eq:t1general},
\ref{eq:conditionD1D2generale}) in the presence of a perfectly
reflecting sphere of radius $R_c$.  We recall that the case $R_{c}>R$
(resp. $R_{c}<R$) is called an entrance (resp. exit) problem.

Using the expressions from Table \ref{tab:ftable}, the coefficients
$\eta_d$ and $X_n$ can be written as
\begin{eqnarray}
\label{eq:eta2_case2}
\eta_{2} &=& \frac{R^2}{4}\left(1-x^2 + \frac{2}{kR}\right) + \frac{R^{2}_{c}}{2} \left(\ln(x) - \frac{1}{kR}\right)  ,  \\
\label{eq:eta3_case2}
\eta_{3} &=& \frac{R^2}{6}\left(1-x^2 + \frac{2}{kR}\right) + \frac{R^{3}_{c}}{3R} \left(1 - \frac{1}{x} - \frac{1}{kR} \right), 
\end{eqnarray}
and
\begin{eqnarray}
\label{eq:Xn2_case2}
X_{n} &=& \frac{x^{n}-1 - \frac{n}{kR} + \left(\frac{R_{c}}{R}\right)^{2n} \left[x^{-n}-1 + \frac{n}{kR}\right]}
{1 + \frac{n}{kR} + \left(\frac{R_{c}}{R}\right)^{2n}\left[1 - \frac{n}{kR}\right]}  
\hskip 18mm (d=2) , \\
\label{eq:Xn3_case2}
X_{n} &=& \frac{x^{n}-1 - \frac{n}{kR} + \frac{n}{n+1} \left(\frac{R_{c}}{R}\right)^{2n+1} 
\left[ x^{-n-1}-1 + \frac{n+1}{kR}\right]}{1 + \frac{n}{kR} + \frac{n}{n+1}\left(\frac{R_{c}}{R}\right)^{2n+1}
\left[1 - \frac{n+1}{kR}\right]}  \quad (d=3) .
\end{eqnarray}

\begin{figure}[h!]
 \centering
\includegraphics[width=0.45\textwidth]{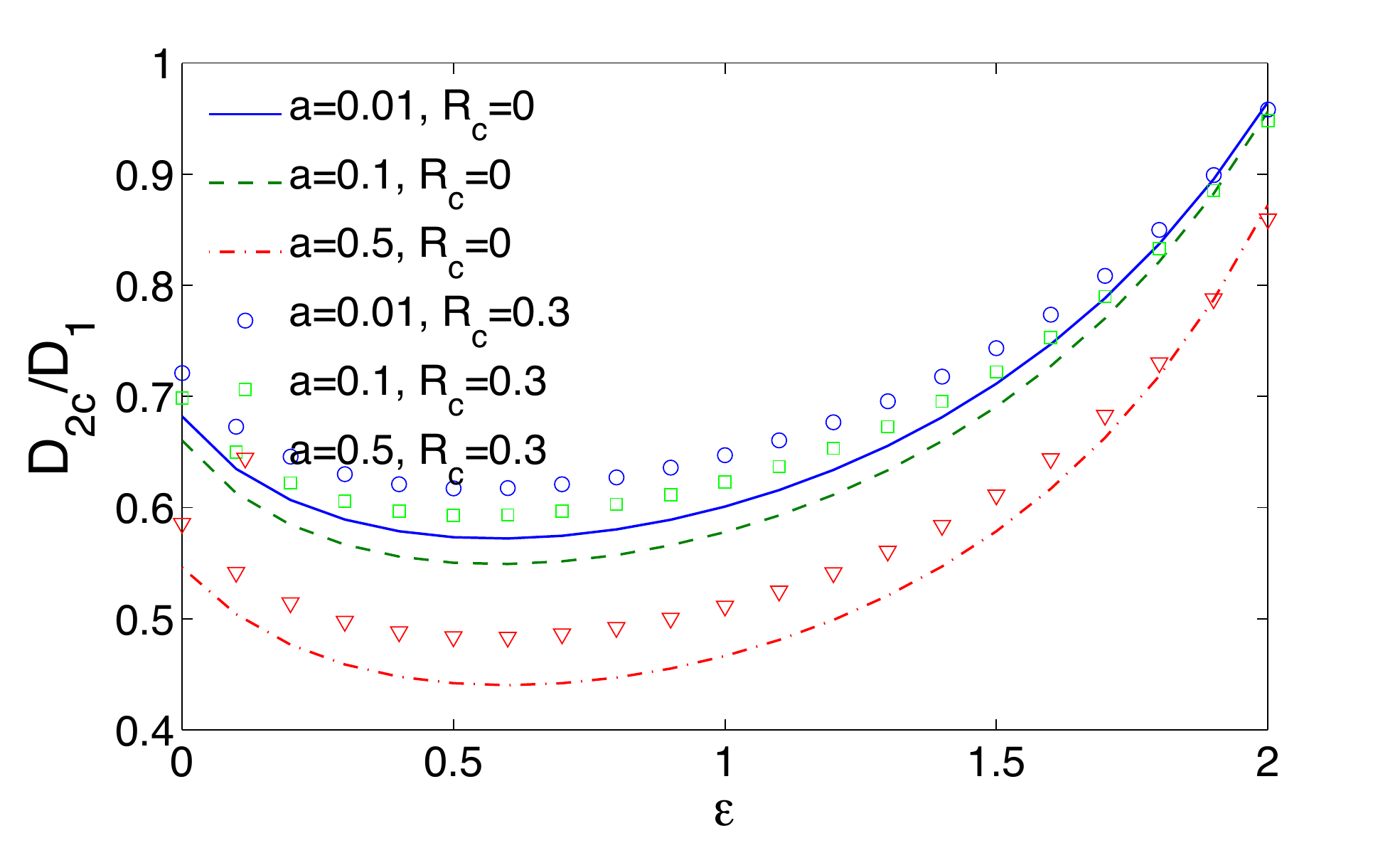}
\includegraphics[width=0.45\textwidth]{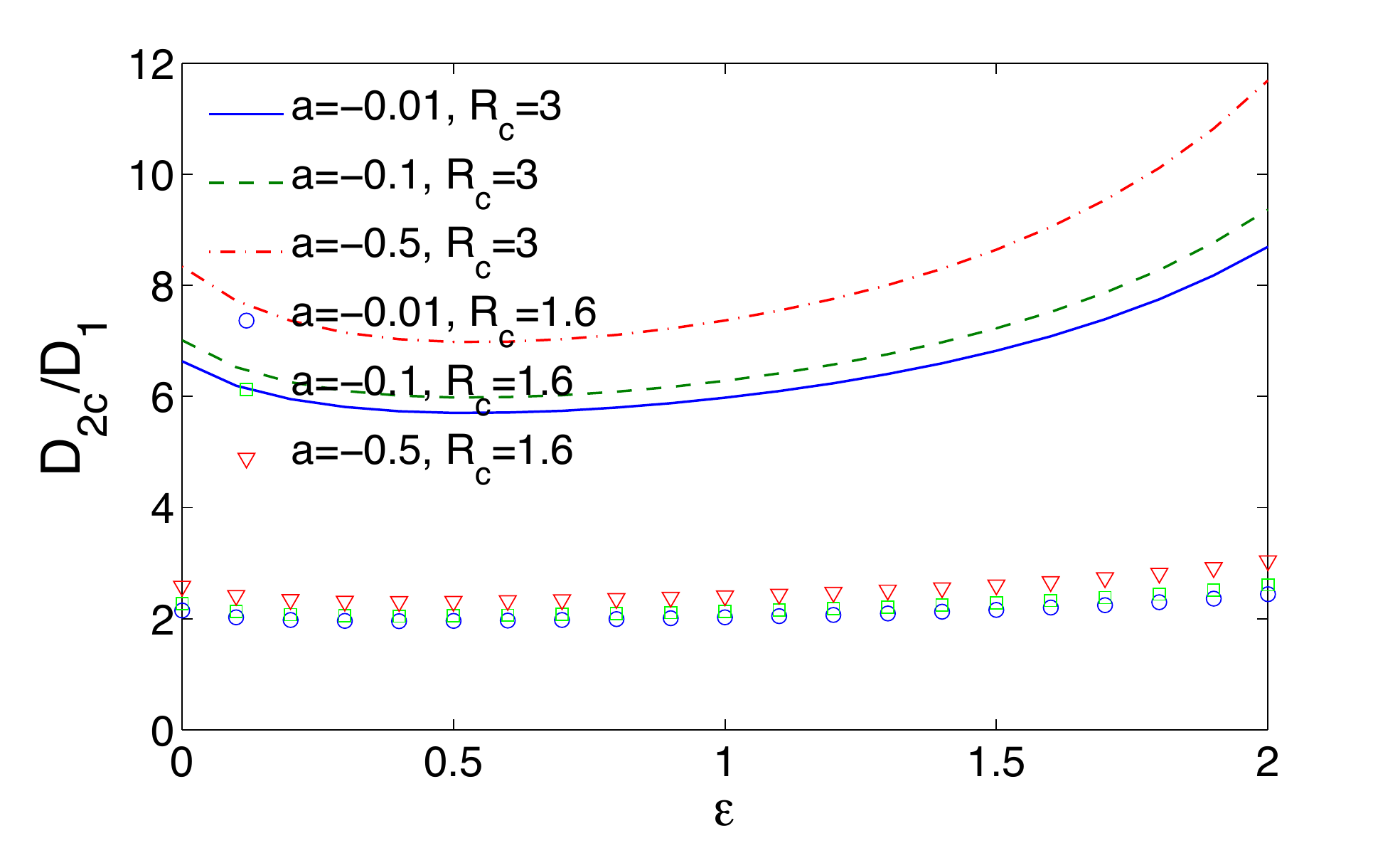}
\caption{
Critical ratio of the bulk-to-surface diffusion coefficients
$D_{2c}/D_{1}$ in 2D, with perfect adsorption and no bias ($k =
\infty$, $V=0$) computed through Eq. (\ref{eq:conditionD1D2generale})
as a function of the target size $\epsilon$ for different values of
$a$ and $R_{c}$: the exit problem ($R_c < 1$) on the left and the
entrance problem ($R_c > 1$) on the right (the truncation size is
$N=200$). }
\label{fig:D1D2epsilon}   
\end{figure}

It is worth noting an interesting dependence of $\langle t_{1}\rangle$
on the radius $R_{c}$ when $R_{c}$ and $a$ are both small.  One finds
in 2D
\begin{eqnarray}
\frac{\partial \langle t_{1}\rangle}{\partial R_{c}}_{\lvert R_{c}=0} = 0, \qquad 
\frac{\partial^2 \langle t_{1}\rangle}{\partial R^{2}_{c}}_{\lvert R_{c}=0} = 
\left(\frac{D_2}{D_{1}} -\frac{\pi^2}{24} \right)\frac{4\lambda a R}{D_{1} D_2},
\end{eqnarray}
and in 3D,
\begin{eqnarray}
\frac{\partial \langle t_{1}\rangle}{\partial R_{c}}_{\lvert R_{c}=0} = 
\frac{\partial^2\langle t_{1}\rangle}{\partial R^{2}_{c}}_{\lvert R_{c}=0} = 0 ,  \qquad
\frac{\partial^3\langle t_{1}\rangle}{\partial R^{3}_{c}}_{\lvert R_{c}=0} = 
\left(\frac{D_2}{D_{1}} - \frac{8}{27} \bigl[2\ln(2/\epsilon) - 1\bigr]\right)
\frac{9\lambda a R}{8 D_{1} D_2}.
\end{eqnarray}
In 2D, as long as $D_{2}/D_1 > \pi^2/24 \approx 0.411$ introducing a
reflecting sphere of small radius $R_{c}$ increases the search time.
This can be understood as follow: increasing $R_{c} \ll R$ increases
the duration of flights between remote and unvisited regions of the
sphere $r=R$, as these flights have to circumvent an obstacle at
$r=R_{c}$.  These long-range flights can reduce the search time only
if they are not too time costly, hence the condition on
$D_{2}>D_{2c}$.  The critical diffusion coefficient $D_{2c}$ increases
with $R_{c} < R$ (Fig. \ref{fig:D1D2epsilon}).

\subsection{Case of a $1/r$ velocity field}

We now examine the case of a radial $1/r$ velocity field $\vec{v}(r)$
characterized by a dimensionless parameter $\mu$:
\begin{equation}
\vec{v}(r) = -\frac{\mu D_{2}}{r^2} \; \vec{r}.
\end{equation}
Substituting the functions $\hat{f}$, $f_0$ and $f_n$ from Table
\ref{tab:ftable} into Eq. (\ref{eq:T}), we can write the
coefficients $\eta_d$ and $X_{n}$ as
\begin{eqnarray}
\label{eq:eta2_case3}
\eta_2  &=& \frac{R^2}{2(2-\mu)} \left[\left(1 - x^2 + \frac{2}{k R}\right) + 
\frac{2}{\mu} \left(\frac{R_c}{R}\right)^{2-\mu}  \left(x^\mu - 1 - \frac{\mu}{k R}\right)\right] , \\
\label{eq:eta3_case3}
\eta_3  &=& \frac{R^2}{2(3-\mu)} \left[\left(1 - x^2 + \frac{2}{k R}\right) + 
\frac{2}{\mu-1} \left(\frac{R_c}{R}\right)^{3-\mu}  \left(x^{\mu-1} - 1 - \frac{\mu-1}{k R}\right) \right] ,  \\
\label{eq:Xn_case3}
X_{n} &=& \frac{\displaystyle x^{\gamma_0 + \gamma_n} - 1 - \frac{\gamma_n + \gamma_0}{kR} + 
\frac{\gamma_n + \gamma_0}{\gamma_n - \gamma_0}
\left(\frac{R_{c}}{R}\right)^{2\gamma_n} \left[x^{\gamma_0 - \gamma_n} - 1 + \frac{\gamma_n - \gamma_0}{kR}\right]}
{\displaystyle 1 + \frac{\gamma_n + \gamma_0}{kR} + \frac{\gamma_n + \gamma_0}{\gamma_n - \gamma_0} \left(\frac{R_{c}}{R}\right)^{2\gamma_n} 
\left[1 - \frac{\gamma_n - \gamma_0}{kR} \right]},
\end{eqnarray}
where
\begin{equation}
\gamma_n = \begin{cases} \sqrt{n^2 + \mu^2/4}    \hskip 25mm (d = 2) \cr
 \sqrt{n(n+1) + (\mu-1)^2/4}    \qquad (d = 3)  \end{cases}  \quad (n \geq 1),
\end{equation}
and $\gamma_0 = \mu/2$ in 2D and $\gamma_0 = (\mu-1)/2$ in 3D.  Note
that in the limit $\mu = 0$, one gets $\gamma_n = n$ ($n \geq 0$) in
2D, and $\gamma_0 = -1/2$ and $\gamma_n = n+1/2$ in 3D, so that the
above results are reduced to the previous case.  The case $\mu = d$
has to be considered separately because $\hat{f}(r) = \frac{r^2}{4}(1
- 2\ln r)$ in both 2D and 3D.

The same expression for $\eta_d$ stands in the cases $\mu = 0$ in 2D and $\mu = 1$ in 3D:
\begin{equation*}
\eta_d  = \frac{R^2}{4} \left(1 - x^2 + \frac{2}{k R}\right) + \frac{R_c^2}{2} \left(\ln x - \frac{1}{k R}\right) . 
\end{equation*}
When $R_c = 0$ and $\mu \geq d$, the MFPT to the
sphere $\eta_d/D_{2}$ diverges, which causes the
critical bulk diffusion coefficient $D_{2c}$ to diverge. 

\begin{figure}
\begin{center}
\includegraphics[width=80mm]{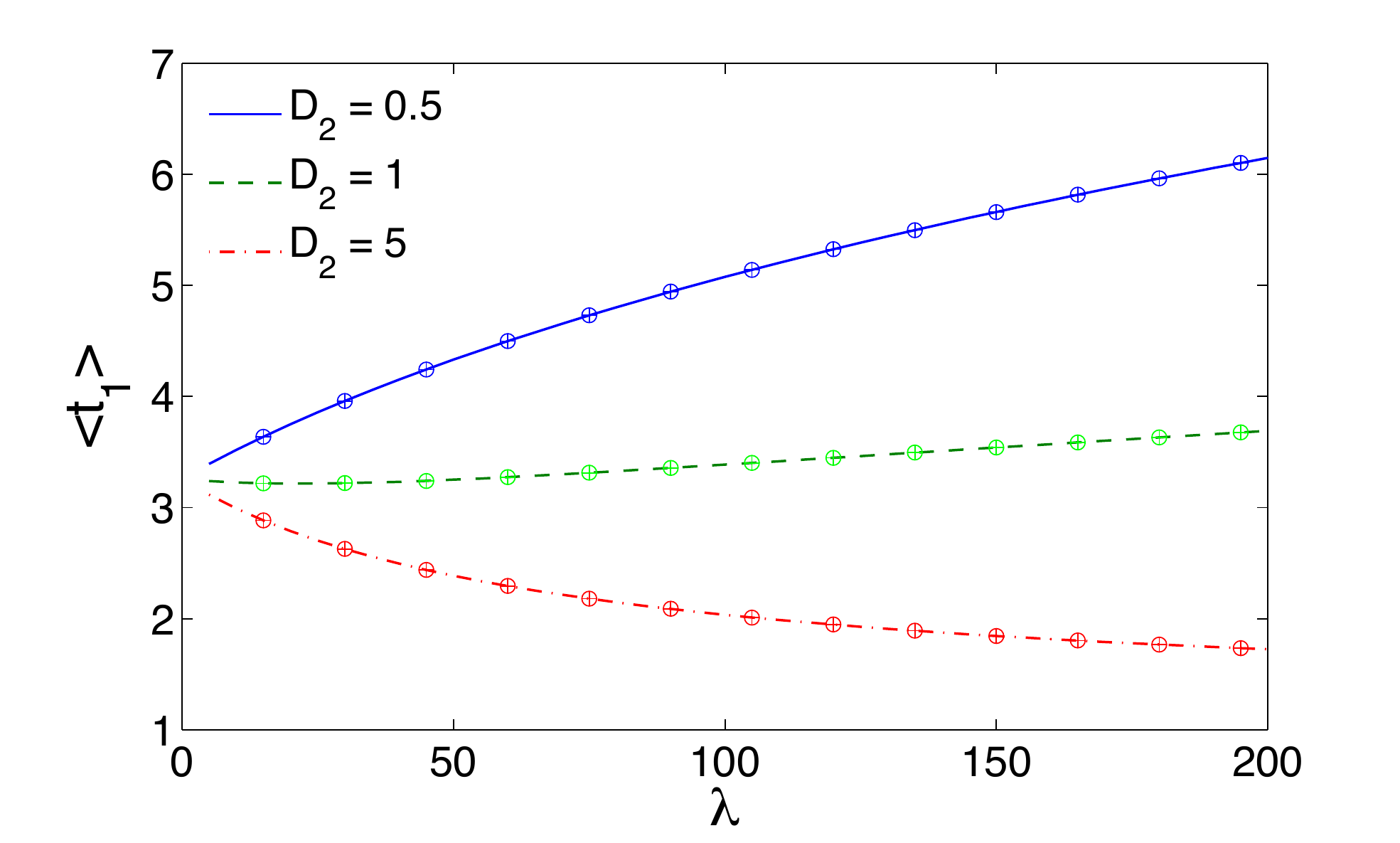}   \includegraphics[width=80mm]{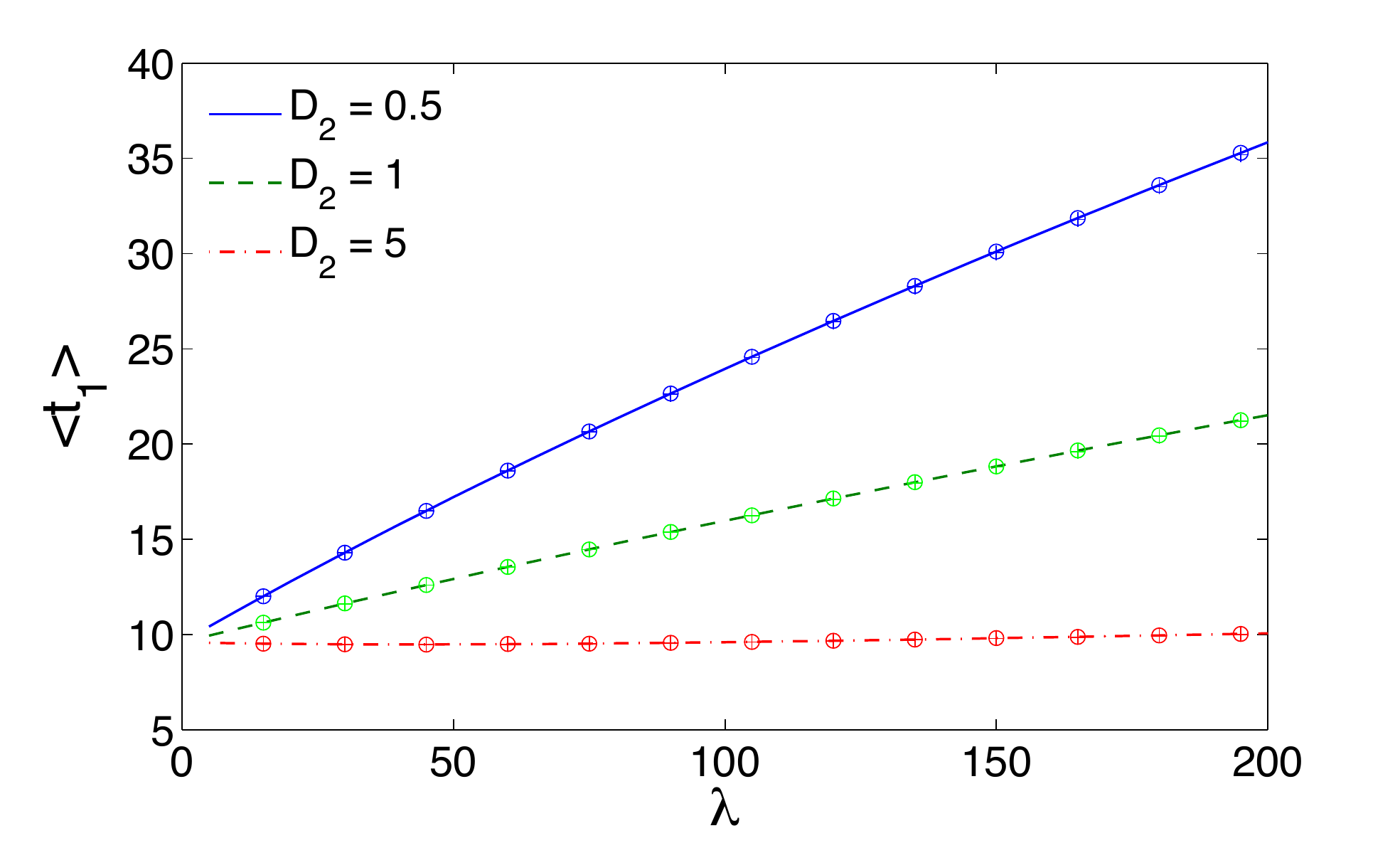}
\includegraphics[width=80mm]{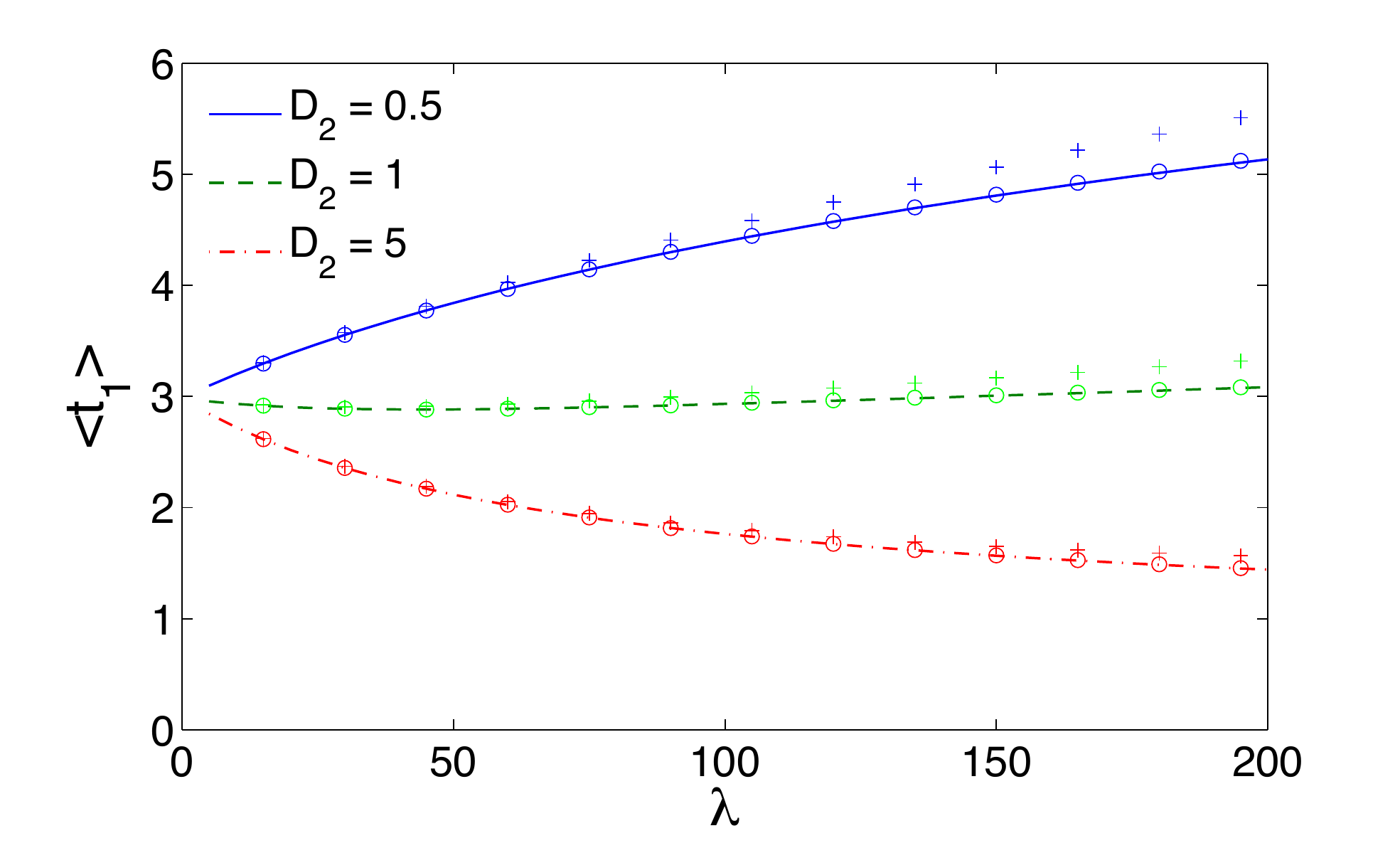}    \includegraphics[width=80mm]{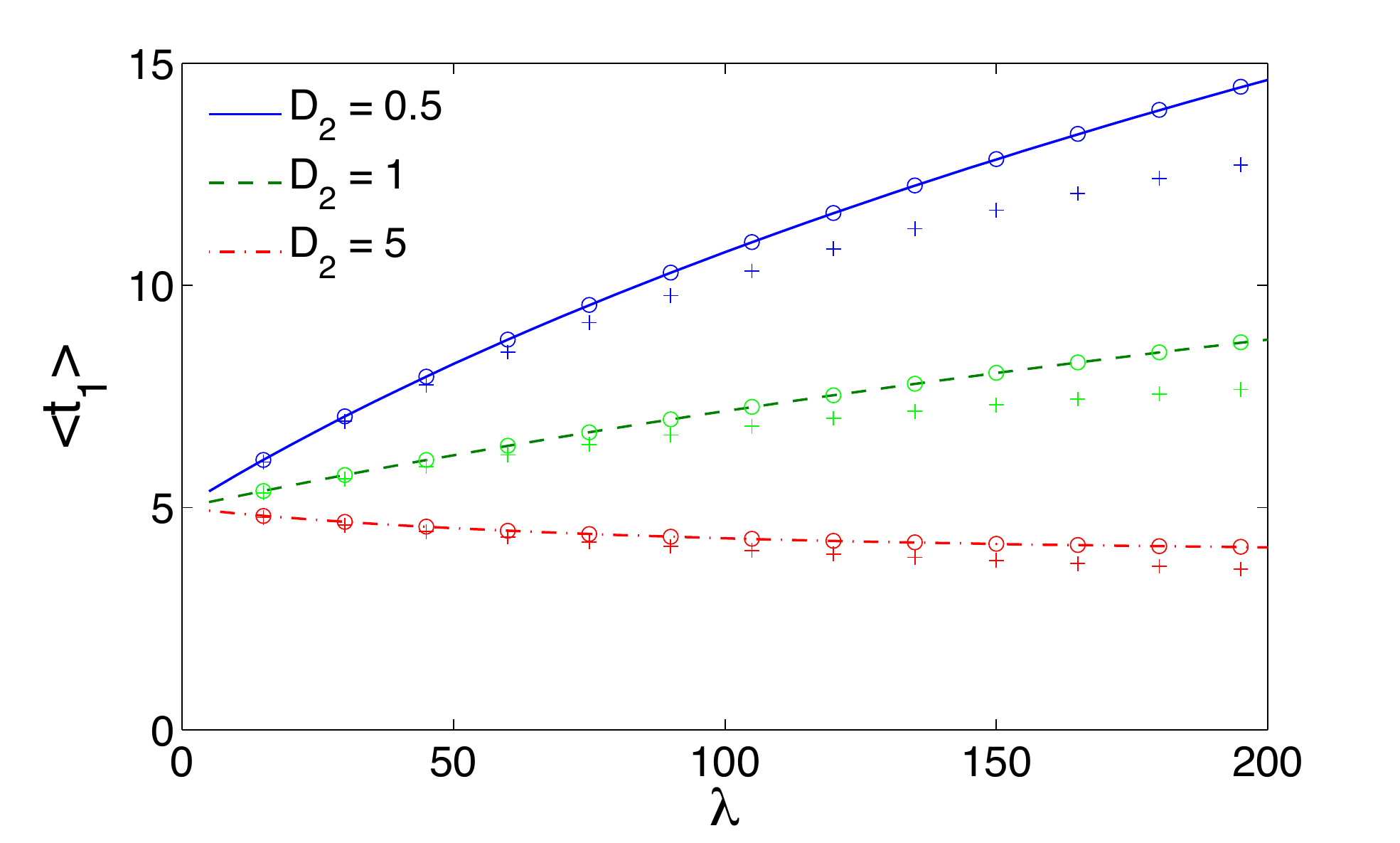}
\end{center}
\caption{
MFPT $\langle t_1\rangle$ as a function of the desorption rate
$\lambda$ in the presence of a $1/r$ velocity field: comparison
between the exact solution (lines), approximate solution (circles) and
perturbative solution (pluses) for 2D with $\mu = 1$ (left) and 3D
with $\mu = 2$ (right), with $\epsilon = 0.01$ (top) and $\epsilon =
0.1$ (bottom).  The other parameters are: $D_1 = 1$, $a = 0.01$, $k =
\infty$, and $D_2$ takes three values $0.5$, $1$ and $5$ (the
truncation size is $N = 200$). }
\label{fig:case3}    
\end{figure}

\begin{figure}[h]
	\centering
\includegraphics[width=0.45\textwidth]{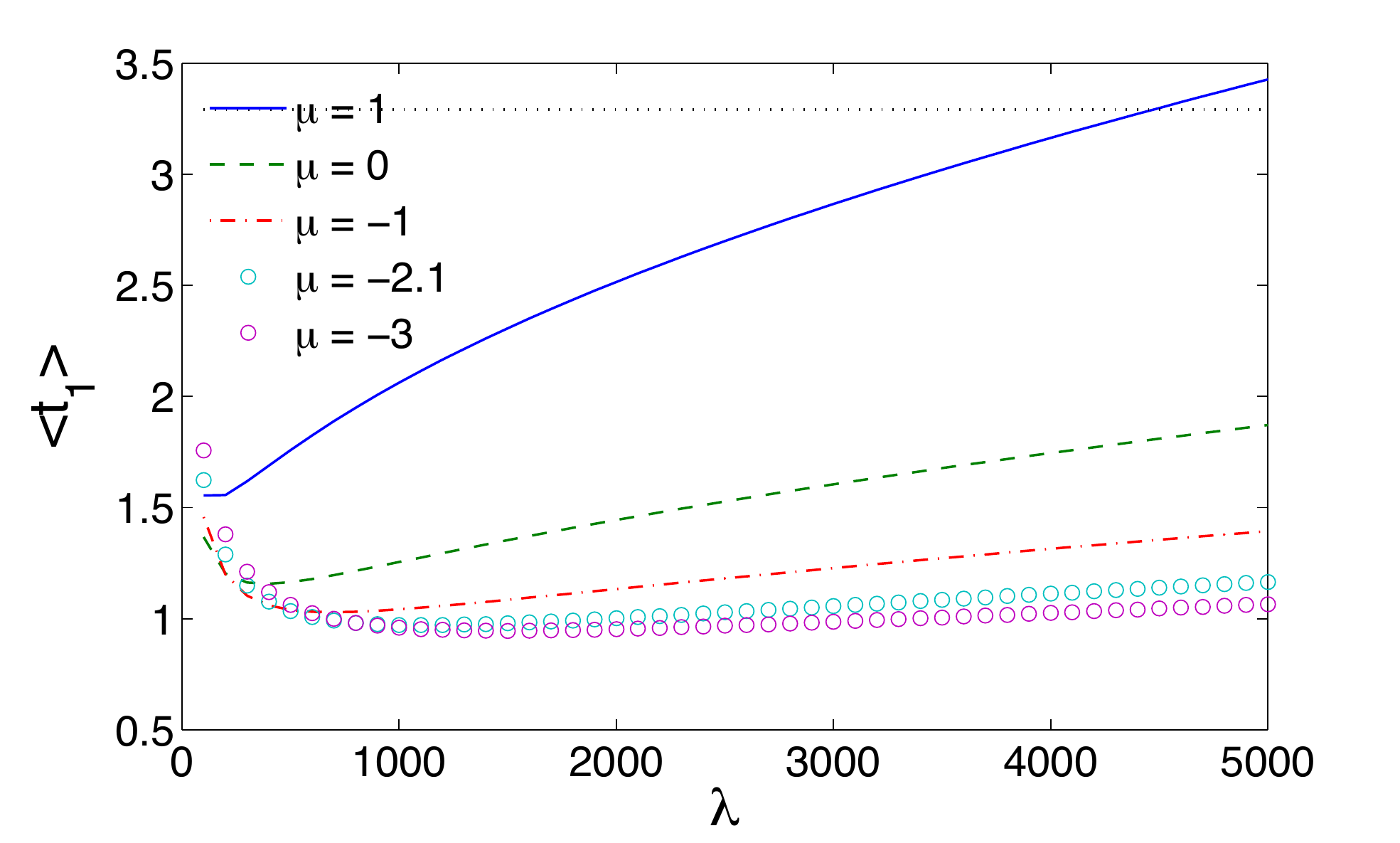}
\includegraphics[width=0.45\textwidth]{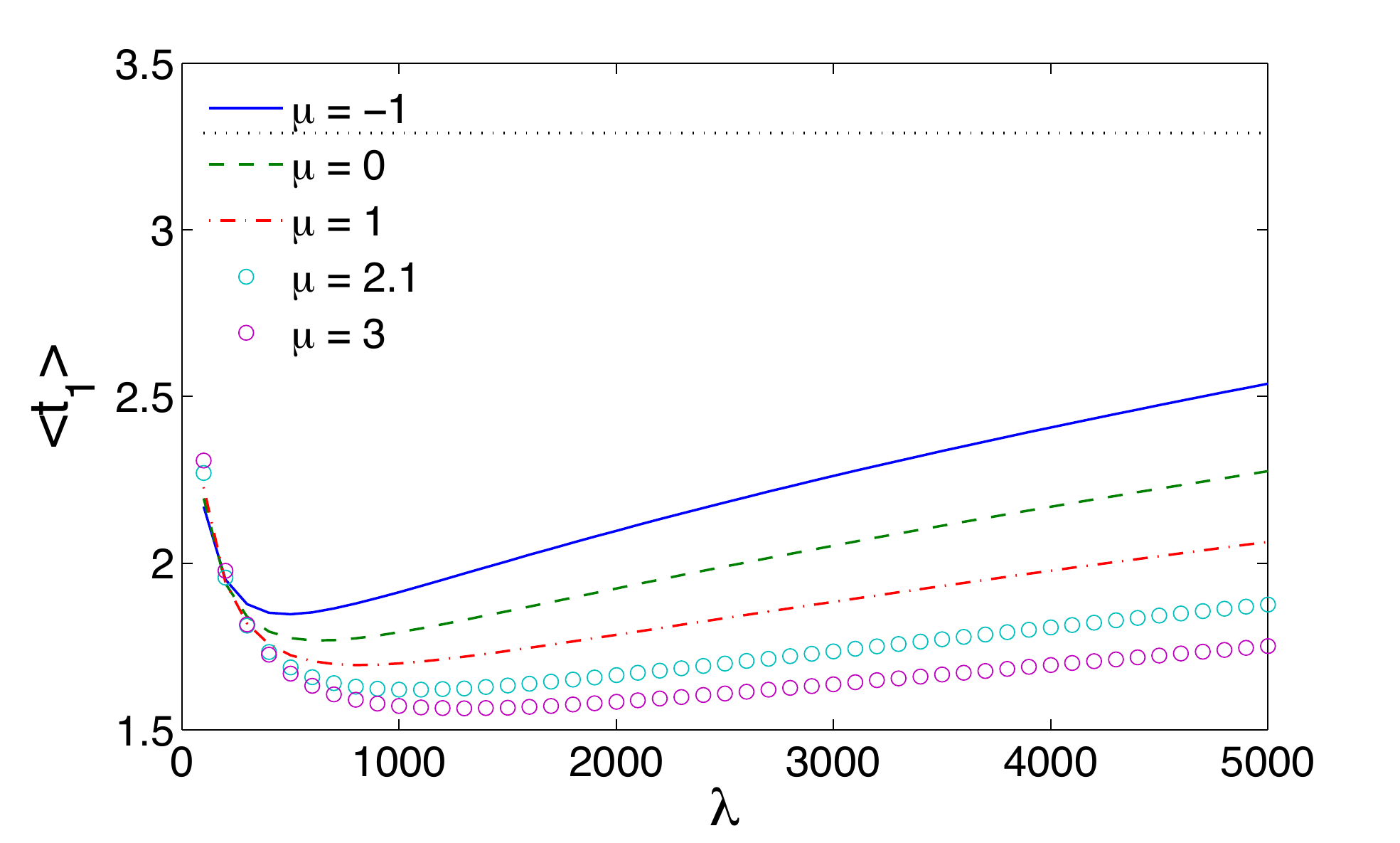}
\caption{
MFPT $\langle t_{1} \rangle$ computed through Eq. (\ref{eq:t1general})
as a function of the desorption rate $\lambda$ for several values of
the drift coefficient for $R_{c}=0$ (left) and $R_{c}=\sqrt{2} > R = 1
$ (right), in 2D.  When $\mu > 0$, the velocity field points towards
the origin, while $\mu < 0$ means that the velocity field points
towards the exterior.  For $R_{c}<R$ (resp. $R_{c}>R$), for a fixed
$\lambda$ the search is on average faster as $\mu$ is more negative
(resp. positive).  Here $d = 2$, $R = 1$, $D_1 = 1$, $D_2 = 5$,
$\epsilon = 0$, $k = \infty$ and $a=0.05$ for $R_{c}=0$ and $a=-0.05$
for $R_{c}=\sqrt{2}$ (the truncation size is $N = 200$). }
\label{fig:t1vslambda(mu)Rapport}   
\end{figure}

\begin{figure}
\begin{center}
\includegraphics[width=80mm]{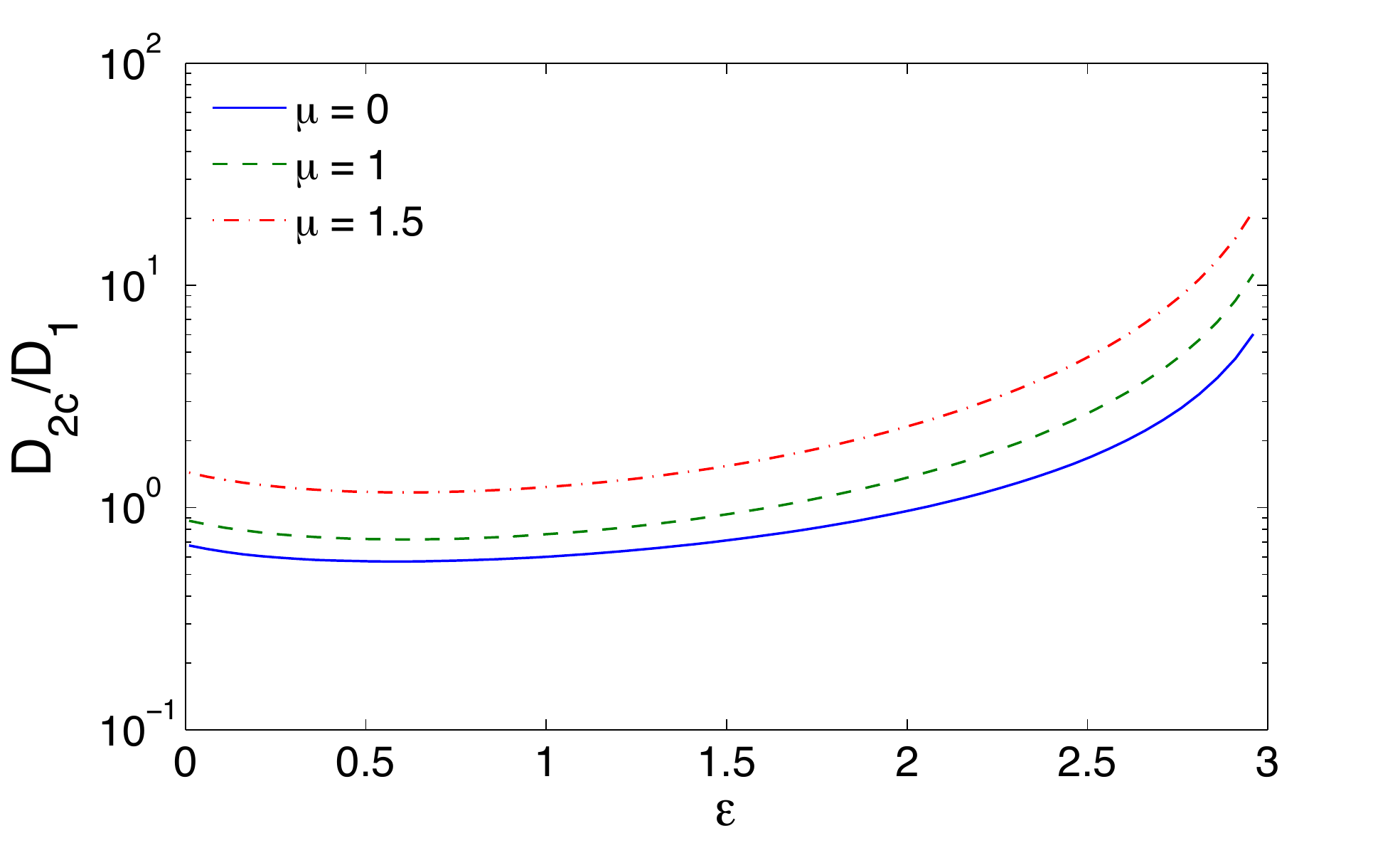}   \includegraphics[width=80mm]{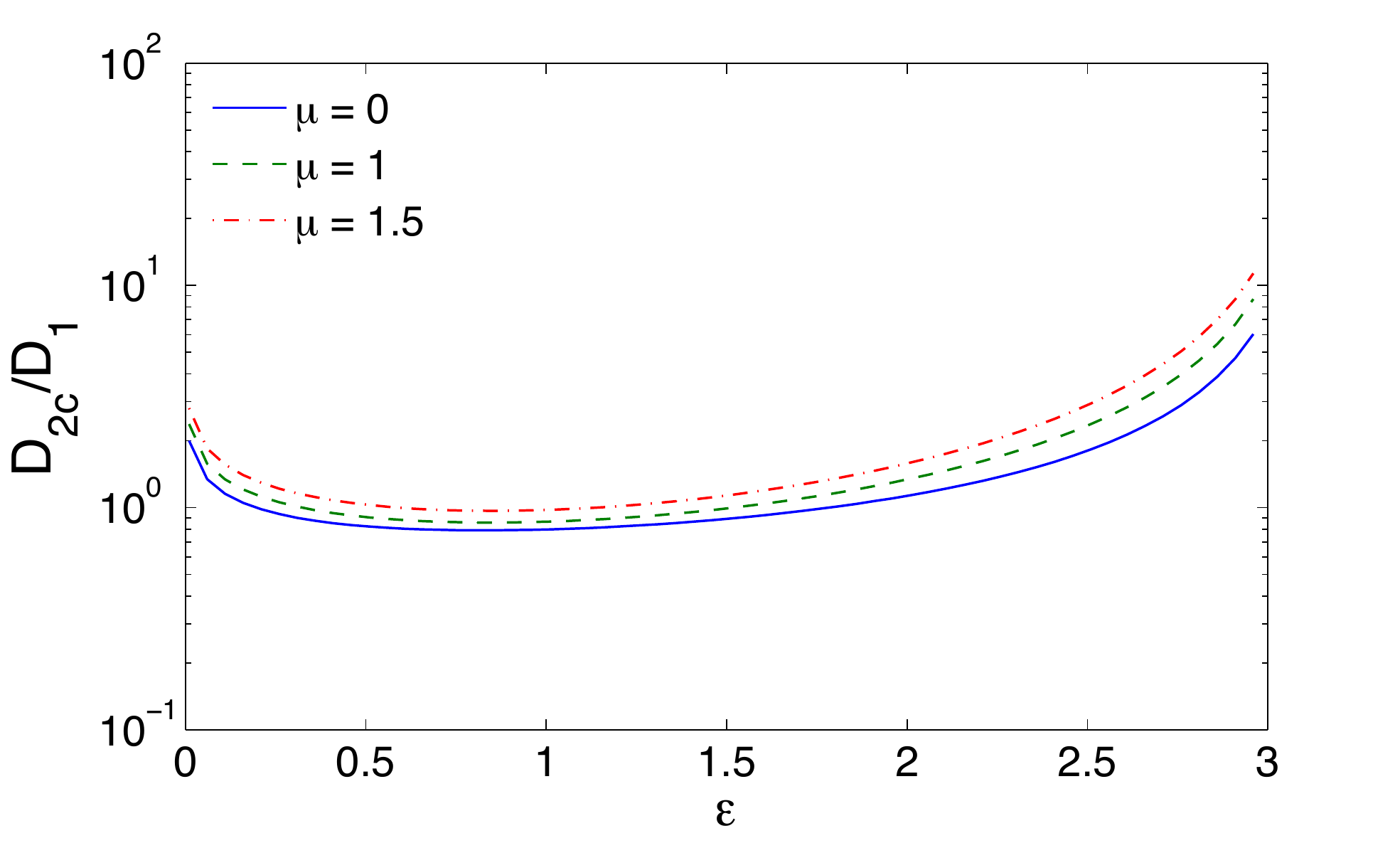}
\end{center}
\caption{
The critical ratio $D_{2c}/D_1$ as a function of the target size
$\epsilon$ in 2D (left) and 3D (right) in the presence of a $1/r$
velocity field with three force intensities: $\mu = 0$ (solid line),
$\mu = 1$ (dashed line) and $\mu = 1.5$ (dash-dotted line).  The other
parameters are: $R = 1$, $a = 0.01$, $R_c = 0$ and $k = \infty$ (the
truncation size is $N = 200$). }
\label{fig:case3_D2cr}    
\end{figure}

Figure \ref{fig:case3} shows the MFPT $\langle t_1\rangle$ as a
function of the desorption rate $\lambda$ in the presence of a $1/r$
velocity field.  As earlier, the exact, approximate and
perturbative solutions are in an excellent agreement for a wide
range of parameters.  Figure \ref{fig:t1vslambda(mu)Rapport} shows a
similar dependence for different field intensities $\mu$ (if $\mu >
0$, the velocity field points towards the origin, while $\mu < 0$
means that the velocity field points towards the exterior).  For
$R_{c}<R$ (resp. $R_{c}>R$), for a fixed $\lambda$ the search is on
average faster as $\mu$ is more negative (resp. positive).  Finally,
in Fig. \ref{fig:case3_D2cr}, the critical ratio of the
bulk-to-surface diffusion coefficients is shown as a function of the
target size, both in two and three dimensions.  The dependence on the
field intensity $\mu$ is stronger in 2D than in 3D.

For $R<R_{c}$, large absolute values of the drift coefficient increase
$\langle t_1\rangle$ and $D_{2c}$ as (i) a strong outward drift
($|\mu| \gg |\mu_{c}|$) diminishes the probability for fast relocation
through the central region; (ii) a strong inward drift ($\mu \gg
|\mu_{c}|$) traps the diffusing molecule in the central region and
increases $\eta_{d}/D_{2}$, the MFPT to the surface $r=R$ after
ejection.

Although we derived the formulas for both 2D and 3D cases, the $1/r$
velocity field is mainly relevant in two dimensions as being a
potential field.  In three dimensions, the potential field exhibits
$1/r^2$ dependence.  This case, as well as many others, can be treated
by our theoretical approach after solving Eqs. (\ref{eq:fhat},
\ref{eq:f0}, \ref{eq:fn_def}) for the functions $f_0(r)$, $\hat{f}(r)$ and
$f_n(r)$.  This is a classical problem in mathematical physics.  For
instance, the aforementioned velocity field $1/r^2$ in three
dimensions involves hypergeometric functions, as shown in Appendix
\ref{sec:Vr2}.

\subsection{ Circular and spherical sectors } \label{sec:sector}

The above approach can also be applied for investigating the MFPTs in
circular and spherical sectors of a given angle $\phi$
(Fig. \ref{fig:case5_nt}).  In most biological situation such as viral
trafficking, $\phi \ll \pi$ but the arguments presented here stand for
arbitrary $\phi$.  For this purpose, the angular basis functions
$V_n(\theta)$ can be rescaled by the factor $\phi/\pi$:
\begin{equation}
\label{eq:Vn_phi}
V_n(\theta)= \begin{cases}
\begin{cases} 1  \hskip 38.5mm (n = 0), \cr
\sqrt{2} \cos(n \theta \pi/\phi) \hskip 17.5mm (n > 0)  \end{cases}  \quad (d = 2) , \cr
\sqrt{2n+1}~ P_n(\cos (\theta \pi/\phi))  \qquad (n \geq 0)  \hskip 9mm (d = 3) ,  \end{cases} 
\end{equation}
and $V_{-n}(\theta) = V_n(\theta)$.  These basis functions satisfy
\begin{eqnarray}
\label{eq:rhophi}
- \Delta_{\theta} V_{n}(\theta) &=& \left(\pi/\phi\right)^2 \rho_{n} V_{n}(\theta)   \qquad (0 \leq \theta \leq \phi, ~n \geq 0), \\
\label{eq:fnphi}
r^2 \left(\Delta_{r} + \frac{v(r)}{D_2}~ \partial_r \right)f_{n}(r) &=& \left(\pi/\phi\right)^2  \rho_{|n|} f_{n}(r) 
\qquad (n\in \Z). 
\end{eqnarray}
As previously, we define two scalar products
\begin{eqnarray*}
(f,g) &\rightarrow& \langle f | g \rangle = \int^{\phi}_{0} f(\theta) g(\theta) d\mu_{d}(\theta), \\
(f,g) &\rightarrow& \langle f | g \rangle_\epsilon = \int^{\phi}_{\epsilon} f(\theta) g(\theta) d\mu_{d}(\theta), 
\end{eqnarray*} 
where $d\mu_{d}(\theta)$ is the measure in polar ($d=2$) or spherical
($d=3$) coordinates for all $\theta \in[0,\phi]$:
\begin{equation*} 
d\mu_{2}(\theta) = \frac{d\theta}{\phi}  \hskip0.5cm{\rm{and}}\hskip0.5cm  d\mu_{3}(\theta) = \frac{\pi}{\phi} \frac{\sin \theta}{2}  d\theta.
\end{equation*}
This modified measure is such that the eigenvectors $V_{n}(\theta)$ are
orthonormal: $\langle V_{n}(\theta) | V_{n'}(\theta)\rangle =  \delta_{nn'}$.

\subsubsection{ Circular sector }

One can easily extend the function $g_\epsilon(\theta)$ for a sector
of angle $\phi$:
\begin{equation}
g_\epsilon(\theta) = \frac12 (\theta - \epsilon)(2\phi - \epsilon - \theta) .
\end{equation}
The direct computation yields
\begin{eqnarray*}
\langle g_\epsilon | 1 \rangle_\epsilon   &=& \frac{(\phi - \epsilon)^3}{3\phi} , \\
\langle g_\epsilon | V_n \rangle_\epsilon &=& - \frac{\phi \sqrt{2}}{\pi^2 n^2}
\left((\phi - \epsilon) \cos(\pi n\epsilon/\phi) + \frac{\phi}{\pi}~ \frac{\sin(\pi n \epsilon/\phi)}{n}\right) ,
\end{eqnarray*}
and
\begin{eqnarray*}
I_{nn} &=& 1 - \frac{\epsilon}{\phi} + \frac{\sin(2\pi n\epsilon/\phi)}{2\pi n}   \hskip 70mm (n\geq 1), \\
I_{nm} &=& \frac{2m^2}{\pi (m^2 - n^2)}
\left(\cos(\pi m\epsilon/\phi) \frac{\sin(\pi n \epsilon/\phi)}{n} - \cos(\pi n\epsilon/\phi) \frac{\sin(\pi m \epsilon/\phi)}{m}\right) 
\quad (m\ne n,~ m,n\geq 1)
\end{eqnarray*}
that generalize formulas from Table \ref{tab:Vtable}.

In order to complete the formulas for search times, one needs to
compute the coefficient $\eta_d$ in Eq. (\ref{eq:eta}) and the
coefficients $X_n$ in Eq. (\ref{eq:Xn}) that incorporate the radial
dependences (e.g., the velocity field $v(r)$ or the partial adsorption
on the boundary).  Since the functions $\hat{f}(r)$ and $f_0(r)$
remain unchanged (see Table \ref{tab:ftable}), the coefficient
$\eta_d$ is given by previous explicit formulas:
Eqs. (\ref{eq:eta2_case2}, \ref{eq:eta3_case2}) with no bias ($V=0$)
and Eqs. (\ref{eq:eta2_case3}, \ref{eq:eta3_case3}) for the velocity
field $1/r$.  In turn, the functions $f_n(r)$ are modified for the
sector because of the prefactor $(\pi/\phi)^2$ in
Eq. (\ref{eq:fnphi}).  For instance, if there is no bias, $f_n(r) =
r^{n\pi/\phi}$, from which
\begin{equation*}
X_n = \frac{\left(x^{n\pi/\phi} - 1 - \frac{n\pi/\phi}{kR}\right) +
(R_c/R)^{2n\pi/\phi} \left(x^{-n\pi/\phi} - 1 +
\frac{n\pi/\phi}{kR}\right)} {1 + \frac{n\pi/\phi}{kR} +
(R_c/R)^{2n\pi/\phi}\left(1 - \frac{n\pi/\phi}{kR}\right)}
\end{equation*}
that extends Eq. (\ref{eq:Xn2_case2}) in 2D.  The case of the velocity
field $1/r$ can be studied in a similar way.

Note that the small $\epsilon$ expansion (\ref{eq:perturb2D}) is
modified as
\begin{equation}
\label{eq:perturb2D_phi}
\frac{\langle t_{1}\rangle}{\omega^2T} = \left(\frac{\phi^2}{3}+2\omega^2 (\phi/\pi)^4 
\sum_{n=1}^\infty \frac{X_n}{n^2(n^2 - \omega^2 X_n)}\right) - \phi \epsilon +
\left(1 - 2\omega^2 (\phi/\pi)^2 \sum_{n=1}^\infty\frac{X_n}{n^2 - \omega^2 X_n}\right)\epsilon^2+O(\epsilon^3).
\end{equation}

\subsubsection{ Spherical sector }

One can also compute the MFPT for a spherical sector of angle $\phi$.
The angular basis functions $V_n(\theta)$ were given in
Eq. (\ref{eq:Vn_phi}), while the function $g_\epsilon(\theta)$
satisfying Eq. (\ref{eq:geps_def}) with $g'_\epsilon(\phi) = 0$ is
\begin{equation}
g_\epsilon(\theta) = \frac{1-\cos\phi}{2} \ln\biggl(\frac{1-\cos\theta}{1-\cos\epsilon}\biggr) +
\frac{1+\cos\phi}{2} \ln\biggl(\frac{1+\cos\theta}{1+\cos\epsilon}\biggr) .
\end{equation}
The integration yields 
\begin{equation}
\langle g_\epsilon | 1\rangle_\epsilon = \frac{(1-\cos\phi)^2}{2}\ln \left(\frac{\sin\phi}{\sin\epsilon}\right) 
+ \frac{1+\cos(\phi)^2}{2}\ln\left(\frac{1+\cos\epsilon}{1+\cos\phi}\right) + \frac{\cos\phi - \cos\epsilon}{2} .
\end{equation}
One also needs to compute the projections 
\begin{equation}
\langle g_\epsilon | V_n \rangle_\epsilon = \sqrt{2n+1} \frac{\pi}{2\phi} \int\limits_{\epsilon}^{\phi} d\theta  \sin \theta ~
g_\epsilon(\theta) P_n(\cos (\theta \pi/\phi)) .
\end{equation}
When $m = \pi/\phi$ is an integer, $\cos (m\theta)$ can be expressed
in powers of $\cos\theta$,
\begin{equation}
\cos(m\theta) = 2^{m-1} [\cos\theta]^m + m\sum\limits_{j=1}^{[m/2]}  (-1)^j \binom{m-2-j}{j-1} \frac{2^{m-2j-1}}{j} [\cos\theta]^{m-2j},
\end{equation}
(here $[m/2]$ is the integer part of $m/2$, and we used the convention
for binomial coefficients that $\binom{n}{0} = 1$ for any $n$) so that
the computation is reduced to the integrals
\begin{equation*}
\begin{split}
J_k &\equiv 2k \int\limits_{\epsilon}^{\phi} d\theta  \sin \theta ~ g_\epsilon(\theta) [\cos(\theta)]^{k-1} \\
& = (1-\cos\phi)([\cos\phi]^k-1) \ln\biggl(\frac{1-\cos\epsilon}{1-\cos\phi}\biggr) 
- (1-\cos\phi)\sum\limits_{j=1}^k \frac{[\cos\epsilon]^j-[\cos\phi]^j}{j} \\
& + (1+\cos\phi)([\cos\phi]^k-(-1)^k) \ln\biggl(\frac{1+\cos\epsilon}{1+\cos\phi}\biggr) 
- (1+\cos\phi)\sum\limits_{j=1}^k (-1)^{k-j}\frac{[\cos\epsilon]^j-[\cos\phi]^j}{j} . \\
\end{split}
\end{equation*}
Using this formula, the projections $\langle g_\epsilon | V_n
\rangle_\epsilon$ can be easily and rapidly computed.  Similarly, one
can proceed with the computation of the matrix elements
$I_\epsilon(n,n')$,
\begin{equation*}
I_\epsilon(n,n') = \sqrt{2n+1}\sqrt{2n'+1} \frac{\pi}{2\phi} \int\limits_{\epsilon}^{\phi} d\theta  \sin \theta ~
P_n(\cos (m \theta)) \bigl[P_{n'}(\cos (m \theta)) - P_{n'}(\cos (m \epsilon))\bigr] ,
\end{equation*}
which are reduced to integrals of polynomials.  When $\pi/\phi$ is not
integer, the above integrals can be computed numerically.

The radial functions $\hat{f}(r)$ and $f_0(r)$ remain unchanged, while
$f_n(r)$ are given in Table \ref{tab:ftable} for the case with no
bias.  The coefficient $\eta_d$ remains unchanged
(cf. Eq. (\ref{eq:eta3_case2})), while the coefficients $X_n$ are
given by Eq. (\ref{eq:Xn_case3}) with $\gamma_0 = -1/2$ and $\gamma_n
= \sqrt{n(n+1)(\pi/\phi)^2 + 1/4}$.  The case of the velocity field
$1/r$ can be studied in a similar way.

\subsubsection{ Multiple targets on the circle } \label{sec:multiple}

\begin{figure}
\begin{center}
\includegraphics[height=50mm]{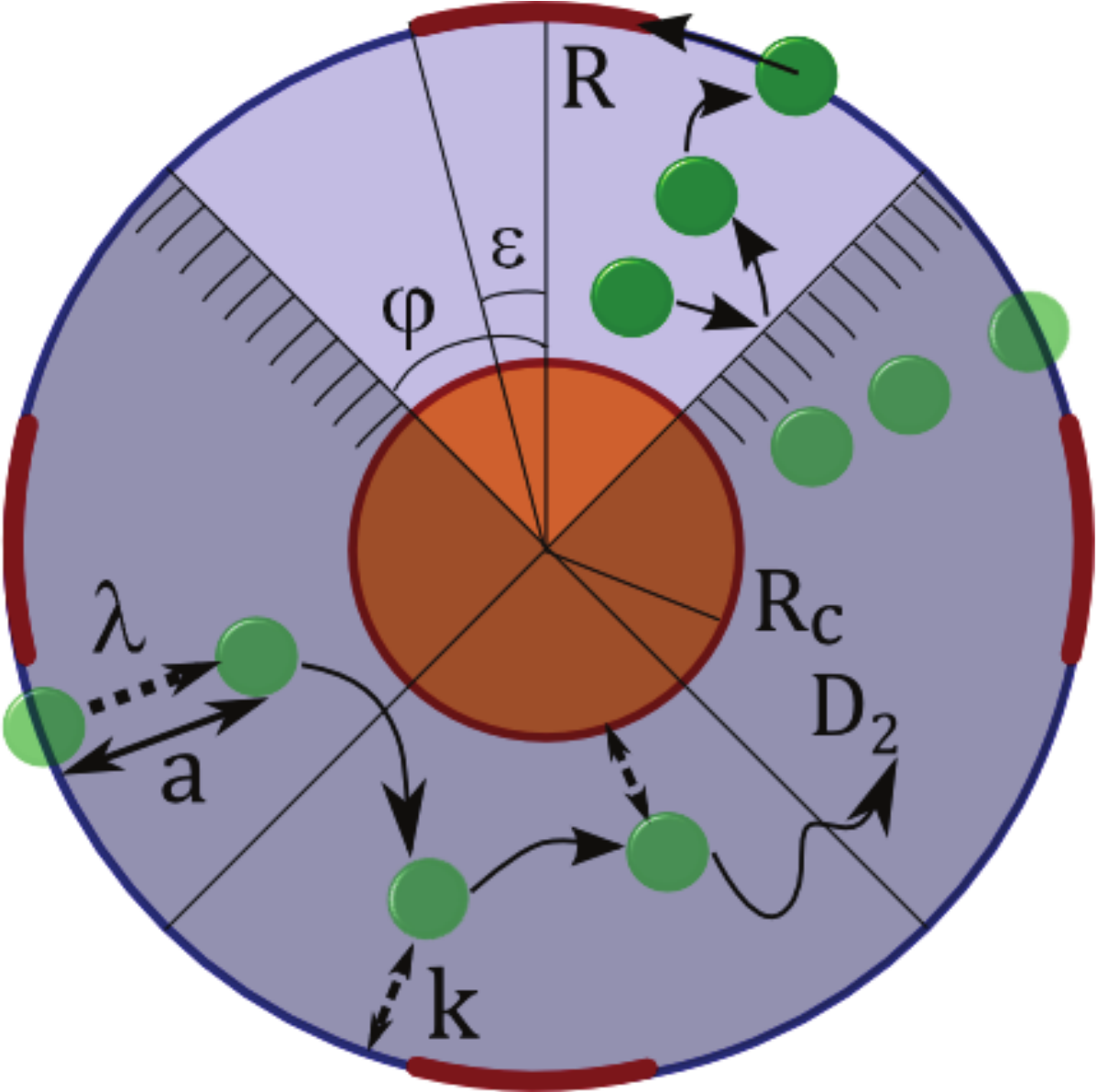} \hskip1cm
\includegraphics[height=50mm]{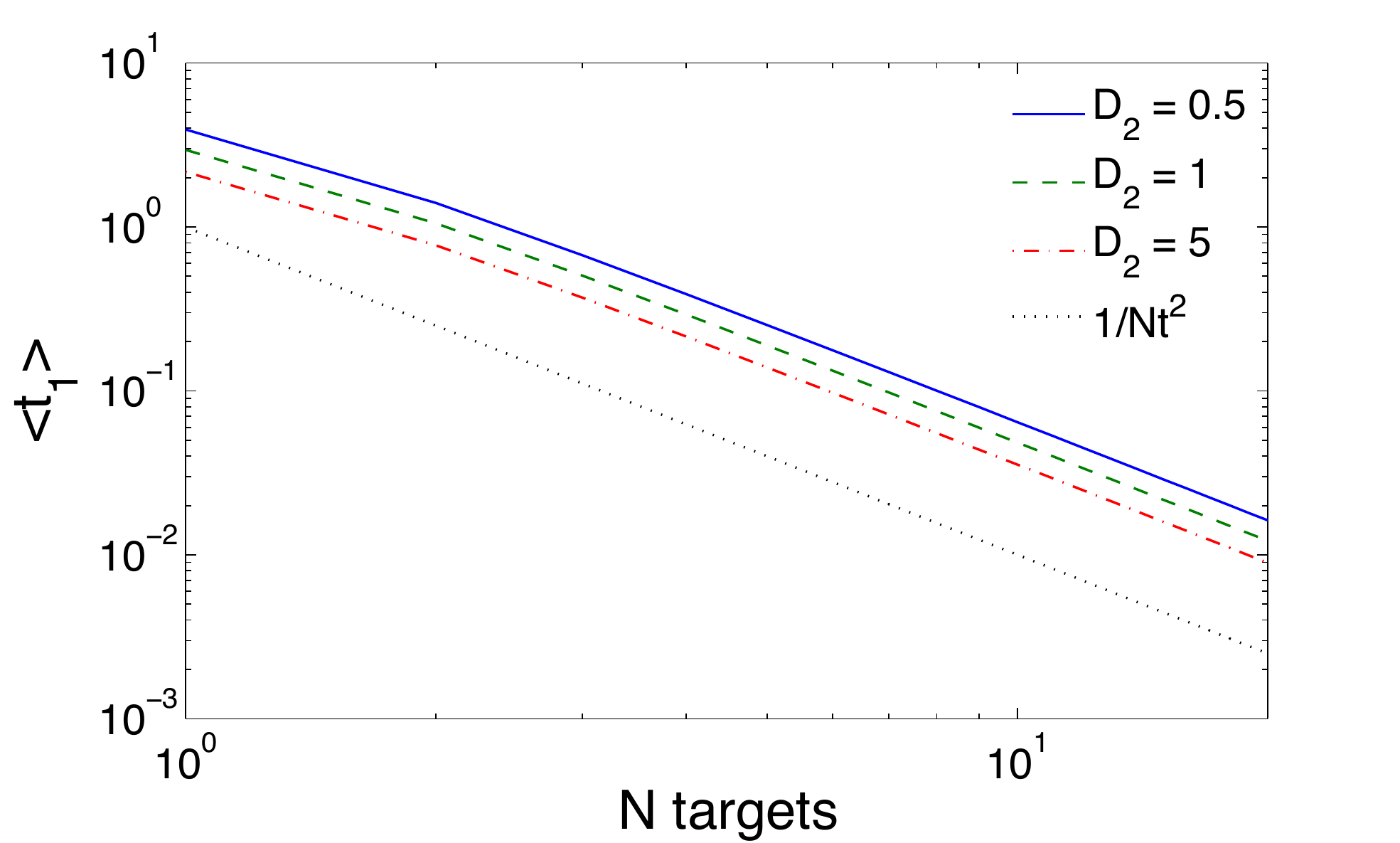}
\end{center}
\caption{\textit{Left} - 
The search problem of four regularly spaced targets can be represented
as a one target search in an angular sector $2\pi/4$ with reflecting
edges.  The shadow green sphere represents the real position of the
molecule in the disk while the solid green sphere represents its
image in the angular sector $2\pi/4$. \textit{Right} - MFPT $\langle
t_1\rangle$ as a function of the number of targets $N_t$ for $\lambda
= 100$, with the total target length $\epsilon_{\rm tot} = 0.01$.
This time decreases as $1/N_t^2$, as one can expect from the limiting
case $\lambda = 0$.  The other parameters are: $R = 1$, $D_1 = 1$, $a
= 0.01$, $R_c = 0$, $k = \infty$, $R_c = 0$, no bias ($V = 0$) and
$D_2$ takes three values $0.5$, $1$ and $5$ (the truncation size is $N
= 200$).  }
\label{fig:case5_nt}   
\end{figure}

The MFPT to reach a target of angular extension $2\epsilon$ in a circular sector of half aperture $\phi = \pi/N_t > \epsilon$ (with integer
$m$) (see Fig. \ref{fig:case5_nt}) can actually be rephrased as the unconditional mean search time of $N_t$ equally spaced
targets of the same size $2\epsilon$ on the circle of radius $R$.
Indeed in 2D, due to the reflection principle for random walks, the time spent to reach any of the $m$ equally spaced targets 
on the circle is equal to the time required to reach a single target within a wedge with reflecting edges at  $\theta = \pm \pi/N_t$.

Figure \ref{fig:case5_nt} shows the MFPT $\langle t_1\rangle$ in 2D as a
function of the number of targets $N_t$, with a fixed total target
length $\epsilon_{\rm tot} = 0.01$.  This time decreases as $1/N_t^2$,
as one can expect from the limiting case $\lambda = 0$.

The same procedure in 3D would be to match the time spent to reach any of $m$ equally spaced target caps of size $2\epsilon<2 \pi/N_t$ on a sphere
with the time required to reach the target cap $\theta \in [-\epsilon,\epsilon]$ of a cone 
with reflecting edges at $\theta = \pm \pi/N_t > \epsilon$ (for all $\phi \in [0,2\pi]$). Although not exact because the volume of a sphere cannot be filled by cones, this procedure is expected to provide an accurate approximation for the unconditional MFPT as soon as the number of targets is sufficiently high. 
For instance, in the case of $60$ equally spaced targets on the sphere, the total
excluded volume (i.e. the volume between cones) represents less than $1\%$ of the total sphere volume. Knowing that the number of membranes or nuclear pores in a cell
usually exceeds $100$ \cite{Lagache:2008a}, the results of Sec. \ref{sec:sector} should to be relevant for cell trafficking studies.

\section{Conclusion}

We have developed a general theoretical approach to investigate searching
of targets on the boundary of a confining medium by surface-mediated
diffusion when the phases of bulk and surface diffusion are
alternating.  This is a significant extension of the previous results
from \cite{Benichou:2010,Benichou:2011a} in order to take into account imperfect
adsorption, the presence of an exterior radial force, multiple
regularly spaced targets and general annulus shapes.
The coupled PDEs for the MFPTs $t_1(\theta)$ and $t_2(r,\theta)$ are
reduced to an integral equation for $t_1(\theta)$ alone whose solution
is then found in a form of Fourier series.  Linear relations for the Fourier
coefficients involve an infinite-dimensional matrix whose inversion
yields an exact but formal solution for the MFPTs.  A finite-size
truncation of this matrix yields a very accurate and rapid numerical
solution of the original problem.  In addition, we propose a fully
explicit approximate solution as well as a perturbative one.  Although both
solutions are derived under the assumption of small targets, the
approximate solution turned out to be remarkably accurate even for
large targets.  We illustrate the practical uses of the theoretical
approach and the properties of the MFPTs by considering in detail
several important examples, for instance diffusion in a velocity $1/r$ field.

The developed approach forms the theoretical ground for a systematic
study of surface-mediated processes which are relevant for chemical
and biochemical reactions in porous catalysts and living cells.  From
the mathematical point of view, the remarkable accuracy of the
approximate solution even beyond the expected range of validity
remains striking and requires further clarifications.  


\acknowledgments
O.B. is supported by the ERC starting Grant FPTOpt- 277998.

\appendix
\section{Boundary condition for the MFPT}
\label{sec:boundary}

We check that Eq. (\ref{eq:adsorption}) giving the discontinuity
relation of the MFPT between the semi-reflecting surface and the bulk
can be derived either from a discrete lattice model or from a standard
forward equation on conditional probabilities
\cite{Redner:2001a,Gardiner:2004}.

\subsection{A discrete lattice approach}
\label{sec:condition}

\begin{figure}[h!]
 \centering
 \includegraphics[width=0.3\textwidth,keepaspectratio=true]{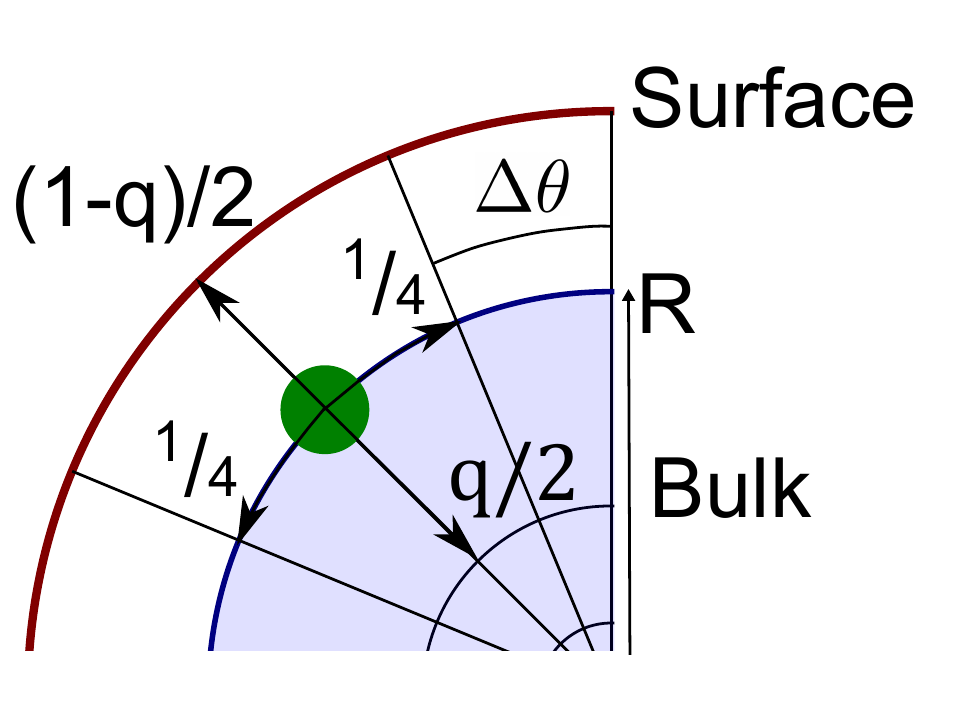}
\caption{
A discrete lattice model.  Regular $\Delta\theta$ slicing imposes
$\Delta t$ and $\Delta r$ to vary over the domain $S$ in order to
maintain an isotropic diffusion constant $D_{2}$.  Only one quadrant
is presented.}
\label{fig:Schemadiscret}
\end{figure}

Let us first consider a 2D geometry in which the bulk and surface
states are two lattices with radial and angular steps $\Delta r$ and
$\Delta \theta$.  The circular geometry imposes the relation on the
radial and angular steps in the bulk at the radius $r$: $\Delta r (r)
= r \Delta \theta $.  At each time step $\Delta t$, the molecule moves
to one of its closest neighboring sites.  The value of the time step
is adjusted according to the position of the molecule:
\begin{equation*}
\Delta t (i,r) = \sqrt{r \Delta \theta} / D_{i} ,
\end{equation*}
where $i=1$ for the molecule on the adsorbing surface and $i=2$ for
the molecule in the bulk.  This choice maintains in the continuous
limit a spatially constant value for the diffusion coefficient
$D_{2}$.  At $r=R$, a molecule may either (i) get reflected to
$r=R-\Delta r$ with probability $q/2$, (ii) get adsorbed onto the
surface with probability $(1-q)/2$, (iii) move along the angular
direction, with probability $1/2$ (see Fig. \ref{fig:Schemadiscret}).
Let the random variable $\tau_2(r,\theta)$ (resp. $\tau_1(\theta)$)
denote the first passage time (FPT) for a molecule initially in the
bulk at $(r,\theta)$ (resp., on the surface at $\theta$).  The
probability for $\tau_2(r,\theta)$ to be $t = m \Delta t$ ($m \in
\mathbb{N}$), is equal to an average of the probabilities of the FPT
from neighboring sites to be $(m-1) \Delta t$:
\begin{eqnarray}
\label{eq:pbords}
\P\biggl\{\tau_{2}(R,\theta)=m \Delta t\biggr\} &=& \frac{q}{2} \P\biggl\{\tau_{2}(R - \Delta r,\theta)=(m-1) \Delta t\biggr\} 
+ \frac{1-q}{2} \P\biggl\{\tau_{1}(\theta)=(m-1)\Delta t\biggr\} \\ 
&+& \frac14 \P\biggl\{\tau_{2}(R,\theta + \Delta \theta) =(m-1) \Delta t\biggr\} + 
\frac14 \P\biggl\{\tau_{2}(R,\theta - \Delta \theta)=(m-1) \Delta t\biggr\} . \nonumber
\end{eqnarray}
The mean FPT in the discrete lattice model is defined as 
\begin{equation} 
\label{eq:tbords}
t_{2}(r,\theta) \equiv \sum^{\infty}_{m=1} m \; \Delta t \;\P\bigr\{\tau_{2}(r,\theta)= m \Delta t\bigr\}.
\end{equation}
Combining this definition with Eq. (\ref{eq:pbords}) leads to
\begin{equation*}
t_{2}(R,\theta) = \frac{q}{2} \; t_{2}(R - \Delta r, \theta) + \frac{1-q}{2} \; t_{1}(\theta) + 
\frac{1}{4}\big( t_{2}(R, \theta + \Delta \theta) + t_{2}(R, \theta - \Delta \theta)\big) + \Delta t.
\end{equation*}
The Taylor expansion of $t_{2}(R,\theta)$ gives
\begin{equation} 
\label{eq:conditiondiscrete}
q \Delta r  \pd{t_{2}}{r}_{\lvert {\bf{r}}=(R,\theta)} + \frac{\Delta \theta^{2}}{2}\pdds{t_{2}}{\theta}_{\lvert {\bf{r}}=(R,\theta)} 
 =  (1-q) \left( t_{1}(\theta) - t_{2}(R,\theta) \right) + O(\Delta r^2,\Delta \theta^{3},\Delta t).
\end{equation}
Following  \cite{Gardiner:2004,Grebenkov:2003}, the adsorption
coefficient is defined as
\begin{equation}
\label{eq:kdefinition} 
k \equiv \frac{1-q}{q}\frac{1}{R \Delta \theta}.
\end{equation}
In the continuous limit, when all $\Delta r,\; \Delta \theta, \;
\Delta t$ tend to $0$ with $\Delta r/\Delta \theta = r $ and $D_{2}
\equiv \Delta r^2/(2\Delta t)$ constant,
Eq. (\ref{eq:conditiondiscrete}) turns out to be expressed in terms of
$k$ only:
\begin{equation} 
\label{eq:conditionbords}
\pd{t_{2}}{r}_{\lvert {\bf{r}}=(R,\theta)} = k \big( t_{1}(\theta) - t_{2}(R,\theta) \big) + o(1).
\end{equation}
Indeed $\Delta t/\Delta r = o(1)$ and $\Delta \theta^{2}/\Delta r =
o(1)$ in this limit.  For a perfectly adsorbing boundary we have
$q=0$, $k\rightarrow \infty$, which is indeed compatible with the
continuity relation $t_{2}(R,\theta) = t_{1}(\theta)$ used in Ref
\cite{Benichou:2011a}.  For a perfectly reflecting boundary $q=1$,
$k\rightarrow 0$ and $\partial_r t_{2}(r,\theta) = 0$ at $r = R$
\cite{Redner:2001a}.  This is indeed the condition that we imposed on
the boundary $r=R_{c}$ in Sec. \ref{sec:general}.

\subsection{Equivalence with a forward boundary condition} 
\label{equivalence}

We now check that the MFPT condition of Eq. (\ref{eq:adsorption}) is
also compatible with the following boundary condition on the
conditional probability \cite{Redner:2001a}
\begin{equation} 
\label{eq:forwardboundary}
\pd{p((r,\theta),t|\bm{x},t')}{r}_{\lvert r=R} = - \; k \; p((r,\theta)|\bm{x},t')_{\lvert r=R},
\end{equation}
where $p(\bm{x},t|\bm{x'},t')$ is the probability for a molecule to be
at $\bm{x}$ at time $t$ provided that the molecule was at $\bm{x'}$ at an
earlier time $t' < t$.  We denote the spatial coordinate $\bm{x} = (r,\theta)$ if the
molecule is in the bulk and $\bm{x} = \theta$ if it is adsorbed on
the surface.

We follow the standard method presented in
\cite{Gardiner:2004}.  The stochastic process under study is Markovian
hence the conditional probabilities satisfy the Chapman-Kolmogorov
equation, with $t>s>t'$,
\begin{equation*} 
 p(\bm{x},t|\bm{x}',t') = \int_{S} d\nu(\bm{y}) \ p(\bm{x},t|\bm{y},s) p(\bm{y},s|\bm{x}',t') 
+ \int^{\pi}_{0} R^{d-1} d\nu(\theta) \ p(\bm{x},t|\theta,s) p(\theta,s|\bm{x}',t'),
\end{equation*}
where $S = (R,R_{c})\times[0,\pi]$ and the measure $d\nu$ is
\begin{equation*} 
d\nu_{2}(r,\theta) = 2 \; r \; dr d\theta, \qquad d\nu_{2}(\theta) =  2 \; d\theta , 
\hskip 10mm d\nu_{3}(r,\theta) = 2 \pi r \sin \theta \; dr d\theta , 
\qquad d\nu_{3}(\theta) = 2 \pi \sin \theta  d\theta .
\end{equation*}
Taking the derivative with respect to the intermediate time $s$ leads
to the expression
\begin{eqnarray} 
\label{eq:chapman2}
 0 = \pd{}{s} p(\bm{x},t|\bm{x}',t') = \int_{S} d\nu(\bm{y}) \ 
\pd{p(\bm{x},t|\bm{y},s)}{s} p(\bm{y},s|\bm{x}',t') + \int_{S} d\nu(\bm{y}) \ p(\bm{x},t|\bm{y},s) 
 \pd{p(\bm{y},s|\bm{x}',t')}{s} \nonumber \\ 
+ \int^{\pi}_{0} d\nu(\theta) \ \pd{p(\bm{\bm{x}},t|\theta,s)}{s} p(\theta,s|\bm{\bm{x}}',t') + \int^{\pi}_{0} d\nu(\theta) \ p(\bm{\bm{x}},t|\theta,s)
  \pd{p(\theta,s|\bm{\bm{x}}',t')}{s} .
\end{eqnarray}
The backward Chapman-Kolmogorov equations read
%
\begin{eqnarray}
\label{eq:Chapapartirde1}
 \pd{ p(\bm{x},t|\theta,s)}{s} &=& - \frac{D_{1}}{R^2} \Delta_{\theta} \ p(\bm{x},t|\theta,s) + 
  \lambda  \biggl\{p(\bm{x},t|\theta,s) - p(\bm{x},t|R-a,\theta,s)\biggr\} , \\	
\label{eq:Chapapartirde2}
 \pd{ p(\bm{x},t|(r,\theta),s)}{s} &=& - D_{2} \ \Delta_{(r,\theta)} \ p(\bm{x},t|(r,\theta),s)  - v(r) \; \nabla p(\bm{x},t|(r,\theta),s) .
\end{eqnarray}
The forward Chapman-Kolmogorov equations are
%
\begin{eqnarray}
\label{eq:Chapvers1}
 \pd{ p(\theta,s|\bm{x}',t')}{s} &=& 
+ \frac{D_{1}}{R^{d}} \Delta_{\theta}p(\theta,s|\bm{x}',t') - \lambda \ p(\theta,s|\bm{x}',t')
- D_{2} \pd{p((r,\theta),s|\bm{x}',t')}{r}_{\lvert r=R} \nonumber \\ && \hskip3cm  + \; v(R) \;  
p((R,\theta),s|\bm{x}',t') , \\
 \pd{ p(\bm{y},s|\bm{x}',t')}{s} &=& D_{2} \ \Delta_{\bm{y}} \ p(\bm{y},s|\bm{x}',t') - \nabla \left( v(r)p(\bm{y},s|\bm{x}',t')\right) \nonumber\\	
\label{eq:Chapvers2}
  && \hskip1.5cm  + \;  \lambda \ \left(\frac{R}{R-a}\right)^{d-1} \; \delta^{d}({\bf{r}}-(R-a,\theta)) \ p(\theta,s|\bm{x}',t') .
\end{eqnarray}
The terms in Eqs. (\ref{eq:Chapvers1}) and (\ref{eq:Chapvers2}) are
justified as follows: (i) $- \lambda \ p(\theta,s|\bm{x}',t')$
corresponds to a constant rate of desorption from the surface to the
bulk; (ii) $-D_{2}\pa_{r} p((r,\theta),s|\bm{x}',t')$ is the flux into
the surface due to diffusion; (iii) $v(R) \;
p((R,\theta),s|\bm{x}',t')$ is the flux into the surface due to the
drift (by convention, $v(R)>0$ for a velocity drift field oriented
towards to the exterior); (iv) $\lambda \
\left[R/(R-a)\right]^{d-1} \; \delta^{d}({\bf{r}}-(R-a,\theta)) \
p(\theta,s|\bm{x}',t')$ corresponds to the flux into the bulk due to
the desorption from the surface and the ejection at a distance $a$
($\delta$ being the Dirac delta function).

For convenience, we will use the shorthand notations
$p(\bm{y},s|\bm{x}',t') \equiv p(\bm{y})$, $p(\bm{x},t|\bm{y},s)
\equiv \bar{p}(\bm{y})$ and $p(\theta,s|\bm{x}',t') \equiv
p(\theta)$.  Substituting the Chapman-Kolmogorov
Eqs. (\ref{eq:Chapapartirde1} -- \ref{eq:Chapvers2}) into
Eq. (\ref{eq:chapman2}) leads to the following equation
\begin{eqnarray*} 
0 &=& \int_{S} d\nu(\bm{y}) \left[- D_{2} \ \Delta_{y \in S} \ \bar{p}(\bm{y})\right] p(\bm{y})  \\ 
&& + \int_{S} d\nu(\bm{y}) \; \bar{p}(\bm{y}) \left[ D_{2} \ \Delta_{y \in S} \ p(\bm{y}) + \lambda \left(\frac{R}{R - a}\right)^{d-1} \;
\delta^{d}({\bf{r}}-(R-a,\theta)) \ p(\theta) \right] \\ 
&& + \int^{\pi}_{0} R^{d-1} \; d\nu(\theta) \ \left[ - \frac{D_{1}}{R^{d}} \Delta_{\theta} \ \bar{p}(\theta) + \lambda
\{\bar{p}(\theta,s) - \bar{p}(R-a,\theta)\} \right] p(\theta,s) \\ 
&& + \int^{\pi}_{0} R^{d-1} \; d\nu(\theta) \ \bar{p}(\theta) \left[ \frac{D_{1}}{R^{d}} \Delta_{\theta}
p(\theta) - \lambda \ p(\theta) - D_{2} \pd{p(r,\theta)}{r}_{\lvert r=R} \right].
\end{eqnarray*}
One notices that
\begin{equation*}
\int_{S} d\nu(\bm{y}) \left(\frac{R}{R-a}\right)^{d-1} \; \delta^{d}({\bf{r}}-(R-a,\theta)) \bar{p}(r,\theta) \ p(\theta) 
+ \int^{\pi}_{0} R^{d-1} \; d\nu(\theta) \bar{p}(R-a,\theta) p(\theta) = 0,
\end{equation*}
and that the four terms proportional to $\lambda$ vanish.  Two terms
with angular Laplacians also cancel each other due to the hermiticity
of the angular diffusion operator:
\begin{equation*}
\int^{\pi}_{0} d\nu(\theta) \ \Delta_{\theta} \ \bar{p}(\theta) p(\bm{y}) - \int^{\pi}_{0} d\nu(\theta) \ 
\Delta_{\theta} \ p(\theta) \bar{p}(\bm{y}) = 0 .
\end{equation*}
The divergence theorem yields the integral over the frontier $\pa S$
of the annulus $S$:
\begin{eqnarray*}
  0 &=& D_{2} \int^{}_{\pa S} d\nu(\theta) \left[ \ \pd{p(r,\theta)}{r}_{\lvert r= R} \bar{p}(r,\theta)
 - \pd{\bar{p}(r,\theta)}{r}_{\lvert r= R} p(r,\theta)
	      - \pd{p(r,\theta)}{r}_{\lvert r= R} \; \bar{p}(\theta,s) \right] \label{eqeq}.
\end{eqnarray*}
This equality can be satisfied only if:
\begin{equation} 
\label{eq:boundary}
 \pd{\bar{p}((r,\theta),s)}{r}_{\lvert r= R} \; p((R,\theta),s) = \pd{p((r,\theta),s)}{r}_{\lvert r= R} 
\biggl[ \bar{p}((r,\theta),s) - \bar{p}(\theta,s) \biggr].
\end{equation}
Inserting the forward boundary condition (\ref{eq:forwardboundary})
into Eq. (\ref{eq:boundary}) gives the boundary condition on the
backward probability distribution
\begin{equation*}
 \pd{p(\bm{x},t|(r,\theta),s)}{r}_{\lvert r= R} = k \biggl[p(\bm{x},t|\theta,s) - p(\bm{x},t|(r,\theta),s)\biggr]_{\lvert r= R}.
\end{equation*}
Integrating over the space and time variables $\bm{x}$ and $t$, we obtain the boundary
condition for the MFPT:
\begin{equation*}
\pd{t_{2}}{r}_{\lvert {\bf{r}}= (R,\theta)} = k \bigl\{ t_{1}(\theta) - t_{2}(R,\theta) \bigr\} \qquad (0\leq \theta \leq \pi),
\end{equation*}
which identifies with Eq. (\ref{eq:adsorption}).

\section{Interpretation of $\eta_{d}/D_{2}$ as a mean first passage time} 
\label{sec:etaMFTP}

We consider the probability density $\Pi (\tp{}|\theta)$ for a
molecule initially at the bulk point $(R-a,\theta)$ to first reach the
surface $r=R$ at the angle $\tp{}$.  The mean duration of this
Brownian path is denoted $t_{c}(\tp{}|\theta)$.

The MFPT $t_{2}(R-a,\theta)$ to reach the target can be expressed as the averaged sum of the MFPT to reach a point $(R,\tp{})$ 
on the surface and the MFPT to reach the target from this point of the surface, 
the probability density for the first hitting point $(R,\tp{})$ being the harmonic measure $\Pi (\tp{}|\theta)$ :
\begin{equation}
 \label{eq:t2alternatif} t_{2}(R-a,\theta) = \int^{\pi}_{0} \left( t_{c}(\tp{}|\theta) + t_{1}(\tp{})  \right) \Pi(\tp{}|\theta) d\mu_d(\tp{}). 
\end{equation}
In 2D and in the general case considered in Sec. \ref{sec:general},
the probability density $\Pi(\tp{}|\theta)$ is
\begin{equation}
 \Pi (\tp{}|\theta) = 1 + 2 \sum^{\infty}_{n=1} (X_{n}+1) \cos( n (\tp{}-\theta)),
\end{equation}
where $X_{n}$ is given by Eq. (\ref{eq:Xn}).
Substitution of this expression in Eq. (\ref{eq:t2alternatif}) leads
to
\begin{equation}
t_{2}(R-a,\theta)  = \langle t_{1}\rangle + \frac{1}{\pi} \int^{\pi}_{0} t_{c}(\tp{}|\theta) \Pi(\tp{}|\theta) d\tp{} 
+ \frac{2}{\pi} \sum^{\infty}_{n=1} (X_{n}+1) \int_0^\pi \cos(n (\tp{}-\theta)) t_{1}(\tp{}) d\tp{}. 
\end{equation}
Identification with Eq. (\ref{eq:t2_final}) gives 
\begin{equation} \label{etareturn}
 \frac{\eta_{d}}{D_2} =  \frac{1}{\pi} \int^{\pi}_{0} t_{c}(\tp{}|\theta) \Pi(\tp{}|\theta) d\tp{}=  
\frac{1}{\pi} \int^{\pi}_{0} t_{c}(\tp{}|0) \Pi(\tp{}|0) d\tp{},
\end{equation}
which identifies $\eta_{d}/D_2$ as the MFPT to the circle
of radius $R$.  In particular, it can be shown that in the 2D case of
Sec. \ref{sec:perfectadsorption}
\begin{equation}
\Pi(\tp{}|\theta) t_{c}(\tp{}|\theta) = \frac{R^2}{4 D_2} \left( 1 - (r/R)^2\right)
 \left( 1 + \sum^{\infty}_{n=1} \frac{(r/R)^n}{2 (1+n)}  \cos(n (\theta-\tp{})) \right).
\end{equation}
One can verify that the substitution of this expression into
Eq. (\ref{etareturn}) leads to the well known result of
Eq. (\ref{eq:etasimple}), $\eta_{2}= R^2(1-x^2)/4$.

The argument leading to Eq. (\ref{etareturn}) can be extended to the 3D case with the following
expression for the probability density:
\begin{equation}
 \Pi (\tp{}|\theta) = 1 + \sum^{\infty}_{n=1} (2n+1) (X_{n}+1) P_{n}(\cos{\tp{}}) P_{n}(\cos{\theta}).
\end{equation}

\section{ Matrix elements $I_\epsilon(n,m)$ in 3D} 
\label{sec:ftable}

The matrix elements $I_\epsilon(n,m)$ in 3D were computed in
\cite{Benichou:2011a}.  An explicit formula for non-diagonal elements
($m\ne n$) is given in Table \ref{tab:Vtable}.  In turn, the diagonal
elements $I_{\epsilon}(n,n)$ can be expressed as
\begin{equation*}
I_\epsilon(n,n) = - P_n(u) \frac{u P_n(u) - P_{n-1}(u)}{n+1} + \frac{F_n(u) + 1}{2n+1},  \hskip 5mm  u = \cos\epsilon ,
\end{equation*}
through the function $F_n(u)$, for which the explicit representation
was derived in \cite{Benichou:2011a}
\begin{equation}
\begin{split}
F_n(u) & = u[P_n^2(u) + 2P_{n-1}^2(u) + ... + 2P_1^2(u) + P_0(u)] \\ 
& - 2P_n(u)P_{n-1}(u) - 2P_{n-1}(u)P_{n-2}(u) - ... - 2P_1(u)P_0(u) + u \\ 
& = \sum\limits_{k=1}^n \bigl[2(u-1)P_k^2(u) + [P_k(u) - P_{k-1}(u)]^2\bigr] - (u-1)P_n^2(u) + (u-1)P_0^2(u) + u .\\
\end{split}
\end{equation}
One can also check that this function satisfies the recurrence relations
\begin{equation}
F_n(u) = F_{n-1}(u) + u[P_n^2(u) + P_{n-1}^2(u)] - 2P_n(u) P_{n-1}(u) , \hskip 5mm  F_0(u) = u . 
\end{equation}
that simplifies its numerical computation.  Note that $F_n(\pm 1) =
F_{n-1}(\pm 1) = ... = \pm 1$.


\section{Case of a $1/r^2$ velocity field}
\label{sec:Vr2}

We now examine the 3D case of a radial $1/r^{2}$ velocity field
$\vec{v}(r)$, which is characterized by the dimensionless parameter
$\mu$:
\begin{equation} 
\vec{v}(r) = -\frac{\mu D_{2} R}{r^3} \; \vec{r}.
\end{equation}
The function $\hat{f}$ is expressed as
\begin{equation}
\hat{f}(r) = -\frac{r^2}{6} - \frac{r R \mu}{6} + \frac{R^2 \mu^2}{6} e^{-\mu R/r}  \Ei(\mu R/r),
\end{equation}
where $\Ei(z)$ is the exponential integral:
\begin{equation*}
\Ei(z) = \int\limits_{-\infty}^z \frac{e^x}{x} dx .
\end{equation*}
The function $f_{0}$ is
\begin{equation}
f_{0}(r) = \frac{1 - e^{-\mu R/r}}{\mu}
\end{equation}
(this particular choice of the additive and multiplicative constants
ensures that $R/r$ is retrieved in the limit $\mu\to 0$).  Radial
functions $f_{n}(r)$ are found as products of powers and confluent
hypergeometric functions $_{1}F_{1}$ of $r$:
\begin{eqnarray}
f_{n}(r) &=& r^{n} \,_{1}F_{1}(-n,\; -2 n,\; -\mu R/r)  \hskip 16mm (n > 0), \\
f_{-n}(r) &=& r^{-n-1} \;  _{1}F_1(n+1,\; 2n+2,\; - \mu R/r) \quad (n > 0) . 
\end{eqnarray}
For $n=0$, this expression is reduced to $e^{-\mu R/r}$.  In the limit
$\mu \to 0$, the above functions reduce to $r^n$ and $r^{-n-1}$ from
the earlier case $\mu = 0$. On the one hand we have
\begin{eqnarray*}
\partial_r f_n(r) &=& n r^{n-1} \,_{1}F_{1}(-n+1,\; -2 n,\; -\mu R/r)  \hskip 16mm (n > 0), \\
\partial_r f_{-n}(r) &=& -(n+1) r^{-n-2} \,_{1}F_{1}(n+2,\; 2n+2,\; -\mu R/r)  \quad (n > 0), 
\end{eqnarray*}
from which
\begin{equation}
\frac{\pa_r f_{n}(r)}{\pa_r f_{-n}(r)} = - \frac{n}{n+1} r^{2n+1} \frac{_{1}F_{1}(-n+1,\; -2n,\; -\mu R/r)}
{_{1}F_{1}(n+2,\; 2n+2,\; -\mu R/r)}.
\end{equation}
On the other hand we have
\begin{eqnarray*}
\partial_r \hat{f}(r) &=& \frac16 \biggl[\frac{(\mu R)^3}{r^2} e^{-\mu R/r} \Ei(\mu R/r) - \frac{(\mu R)^2}{r} - (\mu R) - 2r\biggr] , \\
\partial_r f_0(r) &=& -\frac{R}{r^2} e^{-\mu R/r},
\end{eqnarray*}
from which
\begin{equation}
\frac{\pa_r \hat{f}(r)}{\pa_r f_0(r)} = -\frac{\mu}{6} \biggl[(\mu R)^2 \Ei(\mu R/r) - e^{\mu R/r}\bigl[\mu R r + r^2 - 2r^3/(\mu R)\bigr]\biggr] .
\end{equation}
This last expression is needed to compute the quantities $\eta_{d}$ and $X_{n}$.


\end{document}